\documentclass[iop]{emulateapj}
\usepackage{natbib}
\usepackage{xspace}
\usepackage{booktabs}
\usepackage{amsmath}
\usepackage{color}
\usepackage[utf8]{inputenc}
\usepackage{rotating}

\newcommand{\kepler}{\emph{Kepler}\xspace}
\newcommand{\ktwo}{\emph{K2}\xspace}
\newcommand{\tess}{\emph{TESS}\xspace}

\newcommand{\target}{KIC 2708156\xspace}

\newcommand{\sa}{(\mathbf{S} \cdot \mathbf{X})}

\shorttitle{Multi-wavelength Photometry From Kepler Data}
\shortauthors{Hedges et al.}

\begin{document}


\title{Multi-Wavelength Photometry Derived from Monochromatic Kepler Data}





\author{Christina Hedges}
\affil{Bay Area Environmental Research Institute, P.O. Box 25, Moffett Field, CA 94035, USA}
\affil{NASA Ames Research Center, Moffett Field, CA}

\author{Rodrigo Luger}
\affil{Center for Computational Astrophysics, Flatiron Institute, New York, NY}

\author{Jessie Dotson}
\affil{NASA Ames Research Center, Moffett Field, CA}

\author{Daniel Foreman-Mackey}
\affil{Center for Computational Astrophysics, Flatiron Institute, New York, NY}

\author{Geert Barentsen}
\affil{Bay Area Environmental Research Institute, P.O. Box 25, Moffett Field, CA 94035, USA}
\affil{NASA Ames Research Center, Moffett Field, CA}

\submitted{\today}



\begin{abstract}

The \kepler mission has provided a wealth of data, revealing new insights in time-domain astronomy. However, \kepler's single band-pass has limited studies to a single wavelength.  In this work we build a data-driven, pixel-level model for the Pixel Response Function (PRF) of \kepler targets, modeling the image data from the spacecraft. Our model is sufficiently flexible to capture known detector effects, such as non-linearity, intra-pixel sensitivity variations, and focus change.  In theory, the shape of the \kepler PRF should also be weakly wavelength dependent, due to optical chromatic aberration and wavelength dependent detector response functions.  We are able to identify these predicted shape changes to the PRF using the residuals between \kepler data and our model. In this work, we show that these PRF changes correspond to wavelength variability in \kepler targets using a small sample of eclipsing binaries.   Using our model, we demonstrate that pixel-level light curves of eclipsing binaries show variable eclipse depths, ellipsoidal modulation and limb darkening. These changes at the pixel level are consistent with multi-wavelength photometry. Our work suggests each pixel in the Kepler data of a single target has a different effective wavelength, ranging from $\approx$ 550-750 $nm$. In this proof of concept, we demonstrate our model, and discuss possible use cases for the wavelength dependent Pixel Response Function of \kepler. These use cases include characterizing variable systems, and vetting exoplanet discoveries at the pixel level. The chromatic PRF of \kepler is due to weak wavelength dependence in the optical systems and detector of the telescope, and similar chromatic PRFs are expected in other similar telescopes, notably the NASA TESS telescope.


\end{abstract}

\section{Introduction}
\label{sec:intro}
The \kepler, \ktwo, and \tess missions have demonstrated the power of time series photometry. Each of these missions have performed their observations in a single, wide bandpass, in order to increase signal to noise and reduce complexity of the instrument. This strategy has enabled exquisite time-series photometry. Data from these missions have been used to identify exoplanet candidates (e.g. \cite{DR25}), unlock the inner workings of red giant stars (e.g. \cite{2011Bedding}), unravel the history of our Galaxy (e.g. \cite{2018Silva}), and elucidate the early behavior of supernova (e.g. \cite{2019Dimitriadis}), among many other accomplishments. Beyond single wavelength observations, observing the color and time-series variability of astronomical objects simultaneously has the potential to unlock new insights about astronomical phenomenon.



The NASA \kepler spacecraft has provided photometry of $>$400,000 targets over its 9 year mission life time in a single band pass. In this paper, we discuss how the \kepler data contain a weak wavelength dependency, due to the optical design and detector response. This weak wavelength dependency causes a "chromatic Pixel Response Function", which can be used to reveal multi-wavelength photometry of astronomical targets in the \kepler catalog, by extracting pixel level light curves. Once systematic effects from the detector can be modeled, the weak effect of wavelength dependence in each pixel can be revealed in the residuals. This chromatic information can unlock varied new inference using the wavelength variability of the \kepler sample.


In this paper we demonstrate and validate the wavelength dependence of the \kepler PRF. In Section~\ref{sec:intro} we introduce the characteristics of the \kepler telescope. In Section~\ref{sec:model} we present a data-driven model that can be used to model the \kepler PRF, including \textbf{wavelength-independent, common} instrument systematics.  Using this model we correct instrument systematics in each \kepler pixel independently, leaving any wavelength dependence intact. We use these pixel level light curves to identify weak wavelength dependence in \kepler pixel data. In Section~\ref{sec:EBs} we demonstrate that the wavelength-dependence of the \kepler PRF is detectable by using literature examples of characterized, bright eclipsing binaries. We also present a coarse calibration of the effective wavelengths probed by the \kepler data. In Section \ref{sec:discussion} we discuss applications of this model and how it can be used for a wide range of astrophysical inference, and for vetting planet candidates at the pixel level. We discuss limitations and future improvements on the model. 

\subsection{Characteristics of the \kepler Spacecraft}

During the prime mission (2009-2013), \kepler provided stable, precise time series measurements of $\approx$200,000 targets, each with high precision photometry lasting up to 4 years. These data capture time series variability at extreme precision; Kepler achieved precision better than 30ppm for Kepler band magnitudes less than 12.5 over 6.5 hours \citep{gilliland}. Kepler is capable of observing only at a single wavelength. The \kepler bandpass center is $\approx 640nm$, and has 50\% relative response between 450nm and 825nm. More information on the characteristics of the \kepler telescope can be found in the Kepler Instrument Handbook \citep{kih}, hereafter referred to as KIH. Kepler observations are split into quarters, where each target is observed continuously for 3 months. The spacecraft rotates each season, such that the target falls on a new detector module each season (every fourth quarter the star falls on the same module). During the prime mission the pointing was stable, with pointing better than an arcsecond over the duration of a quarter. This small, long term  drift is caused by velocity aberration \citep[see][]{jenkins2010} and focus change due to changes in spacecraft temperature. These factors cause the position of a target to vary slowly over the course of an observation, characteristically moving $\lesssim$ 1 pixel over the course of a single 3 month quarter. These drifts, coupled with the intrapixel sensitivity variations (see KIH), cause the flux in each pixel to vary slowly over time.

\begin{figure}
    \centering
    \includegraphics[width=0.5\textwidth]{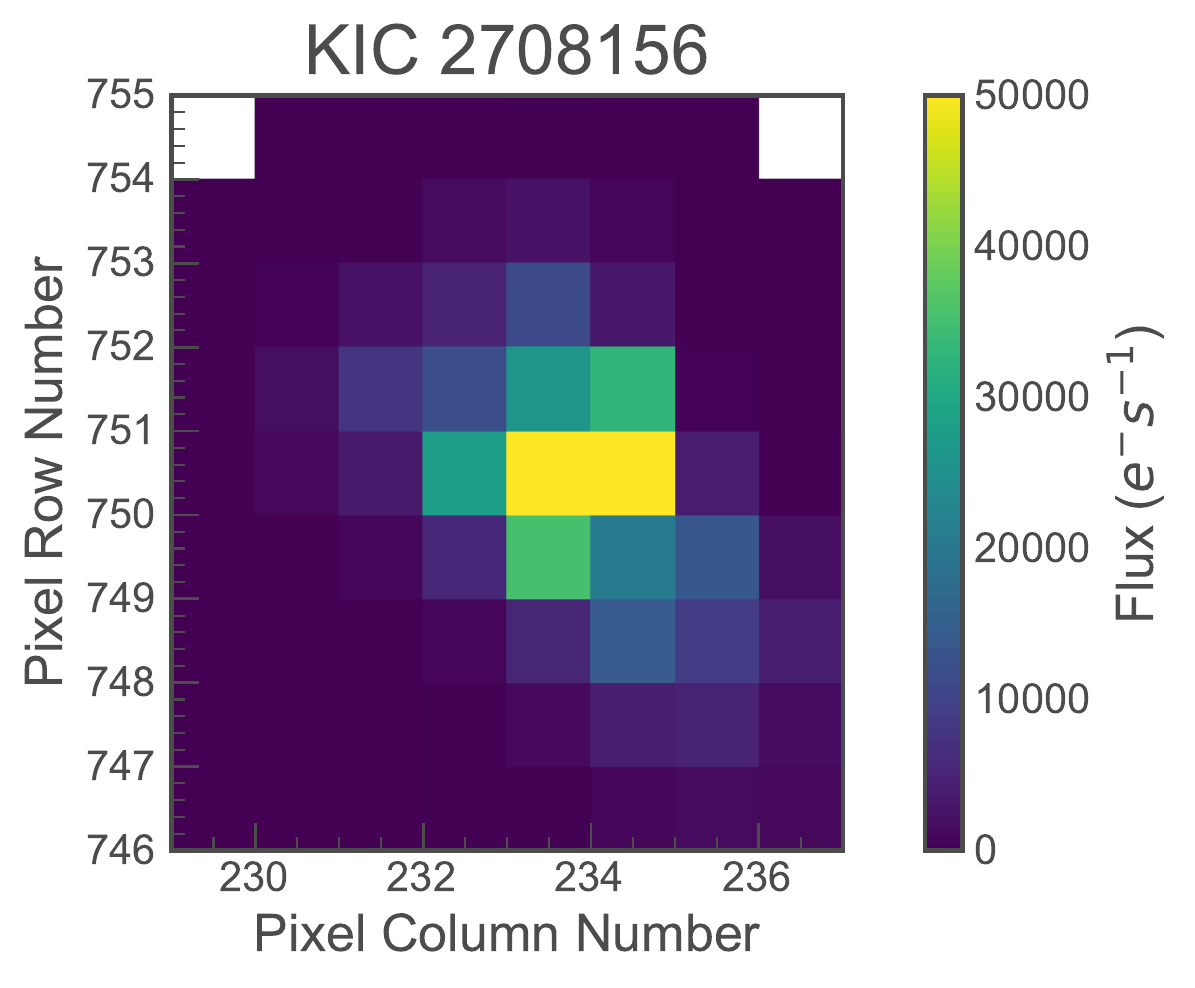}
    \caption{Example of a Target Pixel File (TPF) for target KIC 2708156, which is a bright eclipsing binary from the Kepler field. This figure shows a single frame from the stack of images taken by the telescope. The image contains a single source, but the broad Kepler PSF bleeds flux onto many surrounding pixels. Both of the brightest pixels are above the saturation limit of the detector ($\approx$ 1.5 $\times 10^5$ $e^-/s$), which causes some flux to "bleed" out onto the pixels directly above and below, which is known as a bleed column. See Section~\ref{sec:saturation} for more information on saturation, and bleed columns.}
    \label{fig:tpf_demo}
\end{figure}

The data volume from \kepler was high, owing to its short observing cadence ($\approx$30 minutes in long cadence mode, $\approx$1 minute in short cadence mode). In order to reduce the data volume, the \kepler telescope downlinked data in small image cut outs, centered around targets of interest. Data was delivered in Target Pixel Files (TPFs) which consist of stacks of images typically $\approx 5 \times 5$ pixels across, with one image per 30 minute cadence. An example of a frame from a TPF is shown in Figure \ref{fig:tpf_demo}. TPFs are then converted into light curves by the \kepler Pipeline \citep{pipeline} using Simple Aperture Photometery (SAP). Pixels inside apertures specified by the \kepler Pipeline are summed in order to create a single flux time-series of each target.

When flux from a target goes through the optical system of the telescope, the target flux is spread from a point source into a broad Point Spread Function (PSF). The PSF for \kepler is wider than a single pixel, and so light from the source is registered on multiple pixels (for an example of the theoretical PSF of \kepler see Figure \ref{fig:psf_demo}). The image that is recorded on the detector is known as the Pixel Response Function (PRF), which is the true image combined with the detector sensitivity. For the bright target in Figure \ref{fig:tpf_demo} the PRF covers tens of pixels. 

During the prime mission, systematics in the light curves produced by the mission, (such as focus change and velocity aberration) were corrected by the \kepler pipeline \citep[see Kepler Data Processing Handbook][]{kdph}. The pipeline approach was to use Cotrending Basis Vectors (CBVs) to remove common trends in the light curve sample \citep[see][]{jeff}. In the mission's second phase, known as \ktwo (2013-2018) \citep{howell}, the observations were subject to increased motion noise, owing to the spacecraft pointing in a two-reaction wheel mode. Intrapixel sensitivity variations cause the \ktwo time series to have variability at a level of $\approx$ 0.1\%, and was not explicitly corrected by the \kepler pipeline. The CBVs employed by the pipeline removed common systematics between targets, but owing to the unique sensitivity variations of each pixel, this approach is not able to completely remove the sytematic motion noise. Several groups produced techniques to mitigate the motion noise in the light curves produced by the \ktwo mission, \citep[e.g. See ][]{Vanderburg2014, Luger2016, Oxford}, employing techniques to correct the pixel data local to the source. These works are able to achieve light curves with a precision within $\approx$ 1-4$\times$ that of the prime mission. Since any wavelength dependence in \kepler is predicted to be weak, it is paramount to correct all other known systematics in the \kepler data, such as motion over the intrapixel flat field. In this work, we will use similar techniques to those used to correct \ktwo systematics in \cite{Luger2016} to model systematics from data from the \kepler prime mission at the pixel level, leaving any wavelength dependence intact.

\subsubsection{Saturation}
\label{sec:saturation}
The \kepler detector was designed to ensure that all flux from targets was captured, even in the case that targets would saturate the detector. The saturation limit for the detector is $\approx 1.5 \times 10^5 e^-s^{-1}$ . The \kepler detector is designed such that when a pixel reaches this limit, the flux ``bleeds'' vertically in columns above and below that pixel, creating a ``bleed column.'' This ensures that the total flux of the target can still be measured, even for targets brighter than the saturation limit. This alters the shape of the PRF measured by the detector. Saturation causes unpredictable effects in the pixel level data (e.g. exacerbated systematics from Moire patterns and undershoot), that are discussed in detail in the KIH. Pixels that are at the ends of bleed columns are also highly variable, as flux spills from all the pixels inside the column. In this work, saturated pixels, and pixels at the ends of bleed columns, are removed and not modeled.

\subsection{Known Color Dependence of the \kepler PRF}
\label{sec:color}

There are two known effects that cause a wavelength dependence in the PRF of the \kepler telescope; 1) chromatic aberration from the optical system 2) wavelength dependence of the intrapixel sensitivity. Together, these two effects combine to make the PRF of the instrument weakly color dependent. Below we discuss each of these effects.

\begin{figure}
    \centering
    \includegraphics[width=0.5\textwidth]{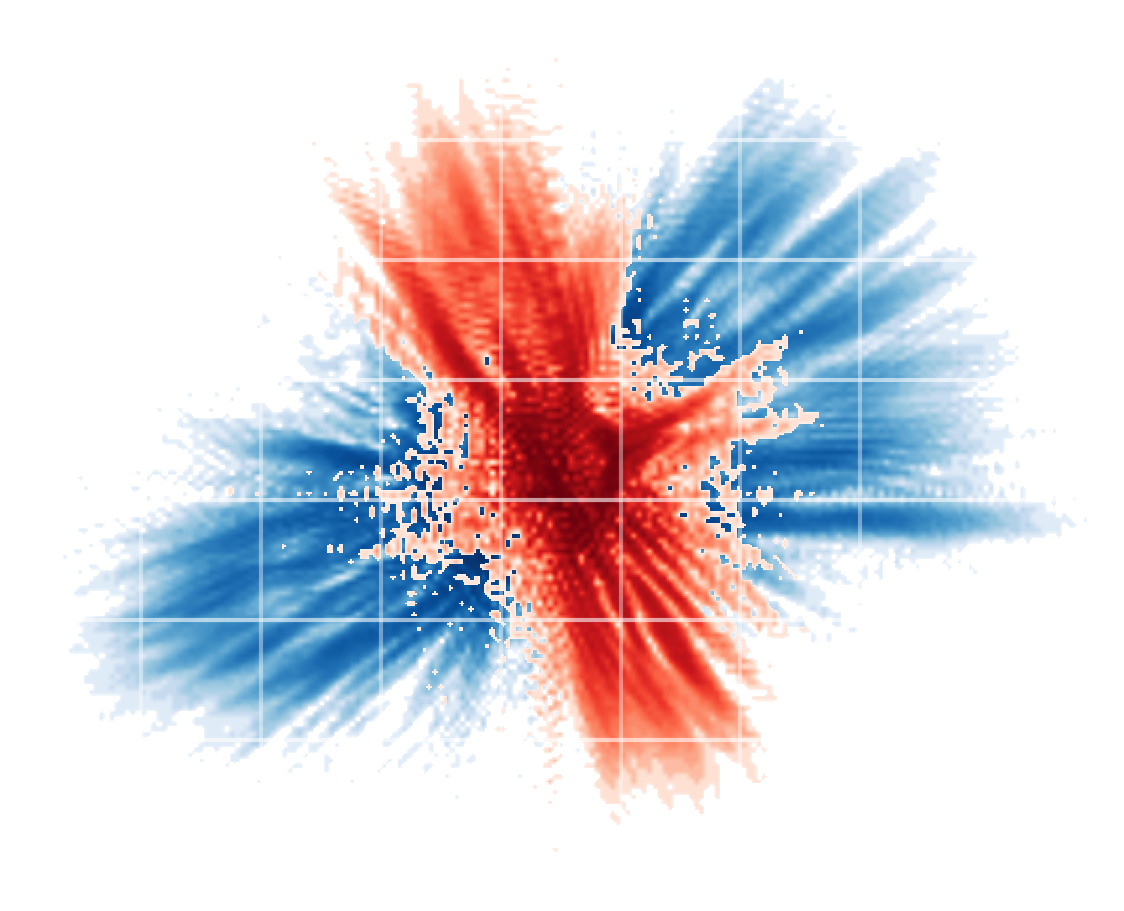}
    \caption{
    Theoretical example of Chromatic Aberration in Kepler PSF, generated by Ball Aerospace using Code V (adapted from the Kepler Instrument Handbook). The PSF is shown for Channel 24.3. The PSF images are histogram-equalization scaled to the same intensity in logarithmic space, and the equivalent width of \kepler pixels is shown by the overlaid grid. The PSF at 430nm is shown in blue, with the PSF at 830nm shown in red. This figure qualitatively demonstrates there is a significant shape change between the PSF at different wavelengths. Note that the true \kepler PSF is integrated across the \kepler bandpass, and so any wavelength variation will be reduced in the real data.}
    \label{fig:psf_demo}
\end{figure}

\subsubsection{PSF Wavelength Dependence}
An example of the theoretical, pre-flight PSF is shown in Figure \ref{fig:psf_demo}, which demonstrates the chromatic aberration. This figure is adapted from the KIH \citep{kih}. The PSF is shown modelled at two different wavelengths at the edges of the \kepler band pass. Figure \ref{fig:psf_demo} shows that, in theory, the PSF shape should vary between these wavelengths. The true PSF is integrated over the \kepler bandpass (which spans $\approx 430 - 830$nm), and so any color variation between the PSFs of blue and red targets is, in reality, difficult to detect. When the PSF is measured (creating a PRF), the shape also varies based on factors such as telescope motion, intrapixel sensitivity, spacecraft temperature, etc. Given these factors, it is difficult to identify these small shape differences in the real data.

The PSF models shown in Figure \ref{fig:psf_demo} are for qualitative purposes. While we can expect some color dependence in the instrument PSF, we do not have an estimate of the magnitude of the effect based on these models. Pre-flight models of the chromatic effects of the \kepler PSF, such as those used to generate the original data behind Figure \ref{fig:psf_demo}, are not publicly available. Such PSF models are likely to have limited use to analyze the color dependence of the \kepler data, as they are based on theoretical models, and do not accurately take into account the detector sensitivities. Also, since these models were generated by the mission on a coarse grid of points around the focal plane, and at a limited set of focus offsets, they would not capture the effect of PSF shape change across a channel.

\subsubsection{Intra-Pixel Sensitivity Wavelength Dependence}

Each pixel in the \kepler detector is not uniformly sensitive. \kepler pixels tend to be most sensitive towards the center, and less sensitive towards the edges \citep[for further details, see][]{Bryson2010}. This effect causes a significant systematic for both \kepler and \ktwo data. \cite{vorobiev} measured the intra-pixel sensitivity using a \kepler flight spare of the detector, and found that the there is significant variation in the pixel response function at 450$nm$, 600$nm$, and 800$nm$. While the same general trend of high sensitivity towards the center of each pixel holds for all wavelengths, the sensitivity drop off at the edges of the pixel is much steeper at shorter wavelengths. A light curve measurement of a target that moves on the detector (either due to spacecraft motion, focus change or velocity aberration) will have a variable flux due to sensitivity variation. Owing only to the wavelength dependence of the sensitivity noted by \cite{vorobiev}, this variability will have a steeper gradient at shorter wavelengths than at longer wavelengths. Similarly, a target that is stationary but undergoes wavelength dependent variability will also have a slightly different response depending on where in the pixel it fell. Even with a completely achromatic PSF, the measured PRF of the instrument will have a wavelength dependence based on this effect alone.

\subsubsection{The PRF}

The true wavelength dependence of the PRF is the combination of both of the effects discussed here. Based on our theoretical understanding of the \kepler PSF and physical characterization of the detector, we can expect a wavelength dependence in the pixel data from \kepler. However, it is not possible based on theory alone to understand the magnitude of these effects. In this work, we develop a model for the instrument systematics of motion and focus change in each pixel, and leave intact any shape variation due to wavelength-dependent variability. Once the wavelength-independent instrument systematics are corrected using our model, we use the residuals to search for shape variation that correlates with astrophysical variability, which we attribute to color dependence.

\section{Data Driven Model For Pixel Systematics}
\label{sec:model}


As discussed in Section \ref{sec:color}, there is no reliable theoretical model for the color dependence of the \kepler PRF. In this work we will instead build a \textbf{data-driven} model for the shape of the PRF, which is flexible enough to capture instrument systematics such as focus change and velcotiy aberration. We use similar techniques to those applied in \cite{Luger2016} and \cite{Vanderburg2014}. Each pixel is modeled independently. We will use this data-driven model to correct the systematics for each pixel individually. We will then identify cadences where there are significant residuals in multiple pixels, indicating there are remaining systematics that are unaccounted for. Where all pixels exhibit a strong deviation from our systematics model, we attribute it to a PRF shape change. Using targets where there is a well known and predictable color change (i.e., eclipsing binaries) we will then show that this PRF shape change is due to wavelength dependent variability. In the following sections we discuss the model for pixel level systematics.

\subsection{General Model}
In this section we describe the data driven model we use to model each pixel light curve, first in a general sense, and then in the specific case of \kepler data in Section \ref{sec:keplermodel}.

In general, we have a time-series measurement of a source, which we label $\mathbf{f}$, which contains both astrophysical time-series and unwanted instrument systematics. In this general section, this could be any flux time-series data; $f$ could be a single pixel time series, or a SAP light curve. Each data point in the time series has some associated error $\sigma_\mathbf{f}$. $\mathbf{K_\mathbf{f}}$ is the covariance matrix for the data.

We label the true, astrophysical flux time-series of the source as $\mathbf{s}$. $\mathbf{s} $ is a vector with $n$ time points, and our goal is to extract the most reliable estimate of $\mathbf{s}$. We have measurements of the source from a telescope, which recorded the source brightness. For this data set, an acceptable model would describe the underlying true astrophysical signal $\mathbf{s}$, and some multiplicative factors and/or additive factors which take into account systematic trends from the telescope. For example, systematic motion over the intrapixel sensitivity function would cause flux to be altered by a multiplicative factor. An unrelated background source would cause an additive offset in the measured flux. The model for our measured signal $\mathbf{f}$ is given as

\begin{equation}
    \label{eq:model}
    \mathbf{f_m} = \mathbf{S} \cdot \mathbf{X} \cdot {\mathbf{w}}
\end{equation}

where $\mathbf{f_m}$ is the model flux, $\mathbf{S}$ is a diagonal matrix where the diagonal values are $\mathbf{s}$, $\mathbf{X}$ is a \emph{design matrix} with predictors for systematics, and $\mathbf{w}$ is a vector of weights for these systematic components.

The \emph{design matrix} $\mathbf{X}$ is a matrix made up of $m$ column vectors, each of length $n$ (the number of time points in the time-series $\mathbf{f}$). Each vector in $\mathbf{X}$ is some predictor of a systematic, such as spacecraft temperature, motion, measured background scattered light etc. These vectors encode our understanding of the systematics. An example of the structure of $\mathbf{X}$ is given in Equation \ref{eq:designmatrix_basic}.

\begin{align}
\label{eq:designmatrix_basic}
\mathbf{X} =  \begin{pmatrix}
              x_{1,1} & x_{1,1} & ... & x_{1,m - 1} & 1\\
              \\
              x_{2,1} & x_{2,2} & ... & x_{2,m - 1} & 1\\
              \\
              ... & ... & ... & ... & ...\\
              \\
              x_{n,1} & x_{n,2} & ... & x_{n,m - 1} & 1\\
              \end{pmatrix}
\end{align}

Note that in the example matrix, the final column of $\mathbf{X}$ is a row of ones. This column allows us to fit an additive offset in the time-series (i.e. to fit the mean level).

In order to calculate the best fitting model, if we know the underlying astrophysical signal $\mathbf{s}$, we must solve for the weights $\mathbf{w}$. In this general case, we assume we know or have a reasonable estimate of $\mathbf{s}$. To solve for $\mathbf{w}$ and find the best fitting model $\mathbf{m}$ we can use linear regression, employing the following relations (under the assumption that our errors are Gaussian)

\begin{equation}
    \label{eq:sigmainv}
    \mathbf{K_\mathbf{w}}^{-1}= \sa^\intercal \cdot \mathbf{K_\mathbf{f}}^{-1} \cdot \sa
\end{equation}

\begin{equation}
    \label{eq:what}
    \mathbf{\hat{w}} = \mathbf{K_w}^{-1} \cdot \left(\sa^\intercal \cdot \mathbf{K_f}^{-1} \cdot \mathbf{f} \right)
\end{equation}

where $\mathbf{\hat{w}}$ is a vector of length $m$ of the mean of the best fitting coefficients and $\mathbf{K_w}$ is the covariance matrix of $\mathbf{w}$ with size $m \times m$. These relations are derived in \cite{themagic}. By rearranging Equation \ref{eq:model} we can now build a systematics model, and find our best estimate of $\mathbf{s}$, the true astrophysical signal.


\subsection{Applying the Model to \kepler}
\label{sec:keplermodel}

\begin{figure}
    \centering
    \includegraphics[width=0.5\textwidth]{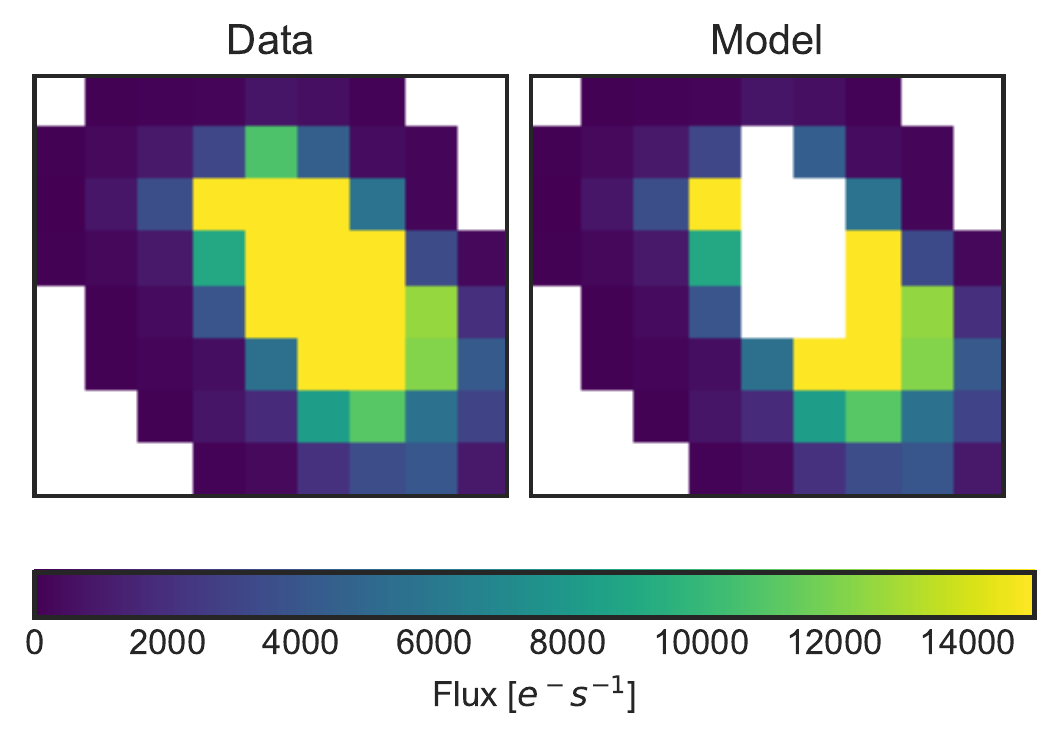}
    \caption{Example of the input data and model built in Section~\ref{sec:keplermodel} for \target. Note that saturated pixels have been omitted in our model, since charge bleed causes the pixel information to be lost. The model image is otherwise indistinguishable from the data, except in cases where there is a slight shape change in the PRF.}
    \label{fig:pix_model_demo}
\end{figure}

In the more realistic case, we have several flux time-series for the source, because the flux has been recorded on several pixels due to the broad PRF. We describe the flux time-series for the $i$th pixel at time $t$ as $\mathbf{f}_{i,t}$. We will assume there are $n$ points in time, $m$ vectors in our \emph{design matrix}, and $l$ pixel time series in our TPF. In this work, we assume there is no covariance between either data points in time, or between each pixel, and our matrix $\mathbf{K_f}$ is a simple diagonal matrix, where the diagonal values are fixed at $\sigma_{\mathbf{f}_i}$ (the error on each flux measurement).
This assumption implies that there are no correlations between either pixels or time points in the dataset, and is undertaken here for simplicity. A more robust analysis might include a $\mathbf{K_f}$ with off-diagonal terms to include covariance in the time-series (encoding covariance between "frames"), and covariance between pixels, (encoding prior knowledge of the "true" PRF shape). In this first proof of concept, we treat each flux measurement as independent.

In this case, our "best guess" for the true underlying signal $\mathbf{s}$ is the Simple Aperture Photometry (SAP) light curve. This light curve is the sum of all pixels within the aperture predetermined by the \kepler pipeline. Our assumption here is that each pixel should contain the same signal as the ``total'' brightness from the source. Said another way, the $\mathbf{s}$ measured in each pixel should be identical (for now, ignoring wavelength dependencies). The SAP flux of a target is simply

\begin{equation}  \mathbf{s} = \sum_{i=0}^{l'}\mathbf{f}_{i,t}\end{equation}

where $\sum_{i=0}^{l'}$ denotes summing the pixel time series inside the optimum aperture specified by the \kepler Pipeline. (Note that $l'$ denotes the pixels inside the aperure, and there are $l$ total pixels in the TPF). If there is no PRF shape variability in our data, the SAP light curve should be identically reflected in every pixel. Because we know there are some unrelated shape and position changes due to focus change, intrapixel sensitivity variations, and velocity aberration, we allow for common systematics between pixels with variable weights using the design matrix $\mathbf{X}$ and weights $\mathbf{w}$.

We build the design matrix $\mathbf{X}$ using the following components:
\begin{itemize}
    \item \textbf{The first two Cotrending Basis Vectors (CBVs) from the \kepler Pipeline.} These are components built by the \kepler pipeline \citep{pipeline}, and reflect common variability across all targets in a single \kepler channel. In our case, they are mostly predictive of focus change. Focus change causes a shape change in the PRF that is common across the entire channel, and has a greater magnitude at the start of quarters (where the spacecraft is changing temperature due to a new pointing). We include these components here, and they are considered a nuisance parameter. If these components are omitted, the PRF shape change due to focus change would not be fit well in our model. We find empirically in our work that the first two Cotrending Basis Vectors are the most predictive of focus change, and sufficiently capture this variability.
    \item \textbf{The pixel column and row positions of the PSF as estimated by the \kepler Pipeline.} These are provided by the TPF in the \texttt{POS\_CORR1} and \texttt{POS\_CORR2} arguments respectively, found in the pipeline processed \texttt{fits} files. We include a 2D, 2nd order polynomial in vectors of row ($\mathbf{r}$) and column ($\mathbf{c}$) position of the target. These vectors are $\{\mathbf{r}, \mathbf{c}, \mathbf{r}^2, \mathbf{c}^2, \mathbf{r}.\mathbf{c},\mathbf{r}^2.\mathbf{c}, \mathbf{r}.\mathbf{c}^2, (\mathbf{r}.\mathbf{c})^2 \}$. This allows us to take into account the motion over the subpixel flat field. For \kepler, the motion takes a simple form and changes gradually, and so a low order 2D polynomial is a reasonable model. For the data taken during the \ktwo mission, this assumption does not hold, and a different model for centroid position is required (see Section~\ref{sec:ktwo}). This polynomial is analogous to the "arclength" used in \cite{Vanderburg2014} to model \ktwo systematics.
    \item \textbf{A higher order polynomial of flux and position.} As shown in other works on the sensitivity of the \kepler pixels, the drop off in sensitivity towards the edge of the pixel is not linear (e.g. see \cite{vorobiev}). As such, we can not model the systematics linearly with $\mathbf{s}$, and we must include higher order terms. To capture this non-linearity, we include columns containing a 1st order polynomial in position $\mathbf{c}$ and $\mathbf{r}$, multiplied by $\mathbf{s}$. The vectors included in the design matrix for this component are $\{\mathbf{s}.\mathbf{r}, \mathbf{s}.\mathbf{c}, \mathbf{s}.\mathbf{c}.\mathbf{r}, \mathbf{s}\}$.
    \item \textbf{A two day B-spline.} In order to capture time variability on short and long time scales, such as from the known \kepler rolling band (short term) and velocity aberration (long term), we include a 3rd order basis-spline in our matrix which has knots every 2 days, this allows us to flexibly capture variability. In this work, where we are concerned with variability due to eclipsing binaries where variability occurs on time scales of hours, a 2 day B-spline is acceptable. However, in other applications, a longer baseline may be more appropriate.
    \item \textbf{A column of ones.} This allows the mean of the time series to be fit by the model. This column allows each pixel to have a different mean. In practice, this enables an offset in each pixel due to background flux from neighboring contaminant stars. This term takes into account the missing offsets from the other polynomial terms given above.


\end{itemize}

An example of the design matrix is shown in Equation \ref{eq:designmatrix}.

\begin{align}
\label{eq:designmatrix}
\mathbf{X} =\begin{pmatrix}
                CBV_{0,0}, & CBV_{1,0}, & r_0, & c_0, & ... & 1\\
                \\
                CBV_{0,1}, & CBV_{1,1}, & r_1, & c_1, & ... & 1 \\
                \\
                CBV_{0,2}, & CBV_{1,2}, & r_2, & c_2, & ... & 1\\
                \\
                ... & ... & ... & ... & ... & ... &\\
                \\
                CBV_{0,N}, & CBV_{1,N}, & r_N, & c_N, & ... & 1\\
                \end{pmatrix}
\end{align}

By utilizing Equations \ref{eq:sigmainv} and \ref{eq:what} we can now find the weights in each pixel $\mathbf{w}_i$ by using the following relations

\begin{equation}
    \label{eq:sigmainv_i}
    \mathbf{K}_{\mathbf{w}_i}^{-1}= \sa^\intercal \cdot \mathbf{K}_{\mathbf{f}_i}^{-1} \cdot \sa
\end{equation}

\begin{equation}
    \label{eq:what_i}
    \hat{\mathbf{w}_i} = \mathbf{K}_{\mathbf{w}_i}^{-1} \cdot \left(\sa^\intercal \cdot \mathbf{K}_{\mathbf{f}_i}^{-1} \cdot \mathbf{f}_i \right)
\end{equation}

where $\hat{\mathbf{w}}_i$ are the best fitting weights in pixel $i$, and is a vector with length $m$, with one coefficient per weight.

Our full model $\mathbf{f_m}$ (given by Equation~\ref{eq:model}) is a matrix of size $n \times l$, where the model has been computed for each time $t$ in each independent pixel $i$. The resultant model is a stack of ``images'' of our data-driven model, including instrument systematics, under the assumption that the astrophysical signal $\mathbf{s}$ is the same in each pixel, i.e. \textbf{a model of the PRF, where the PRF shape may vary only in a way correlated with the CBV components, or with the 2-day basis-spline components.}. An example of the data and the model are shown for \target in Figure~\ref{fig:pix_model_demo}. Using this model, we can now analyze the residuals to see if the PRF shape change does indeed remain constant.

\subsection{Overfitting}

In this work we fit each pixel time series individually, as shown in Equations~\ref{eq:sigmainv_i} and ~\ref{eq:what_i}. We are able to do this only when SNR in each pixel time series is high enough to measure the intrinsic astrophysical variability, (i.e. for bright targets with significant variability). This is the case for bright eclipsing binaries observed with \kepler, but for fainter targets with lower amplitude variability this may not modeled well by the model discussed here. Further investigation is needed to find the limit of the method described here.

Finding $\mathbf{w}_i$ by jointly fitting multiple pixels and multiple targets, taking into account covariances between pixels, would enable us to find a model for the full detector, creating a data-driven PRF model in absolute terms. Such a model would be able to, given an input flux value for a given time and pixel ($\mathbf{f}_{i, t}$) return the estimated flux for a given pixel based on the full context of the detector, accounting for all common systematics. However, to create such a model would require incorporating further information into the model, including the true underlying image (i.e. a sky catalog of the star field, including exact, sub-pixel positions of every target). Such a model is not necessarily easily described by simple, linear combinations of polynomials, and would be computationally intensive to generate. Generating and testing such a model is beyond the scope of this initial work, but would in theory enable this model to be expanded to the full \kepler sample.

\subsection{Model example: Kepler-1b}
\label{sec:exampleplanet}

\begin{figure*}
    \centering
    \includegraphics[width=\textwidth]{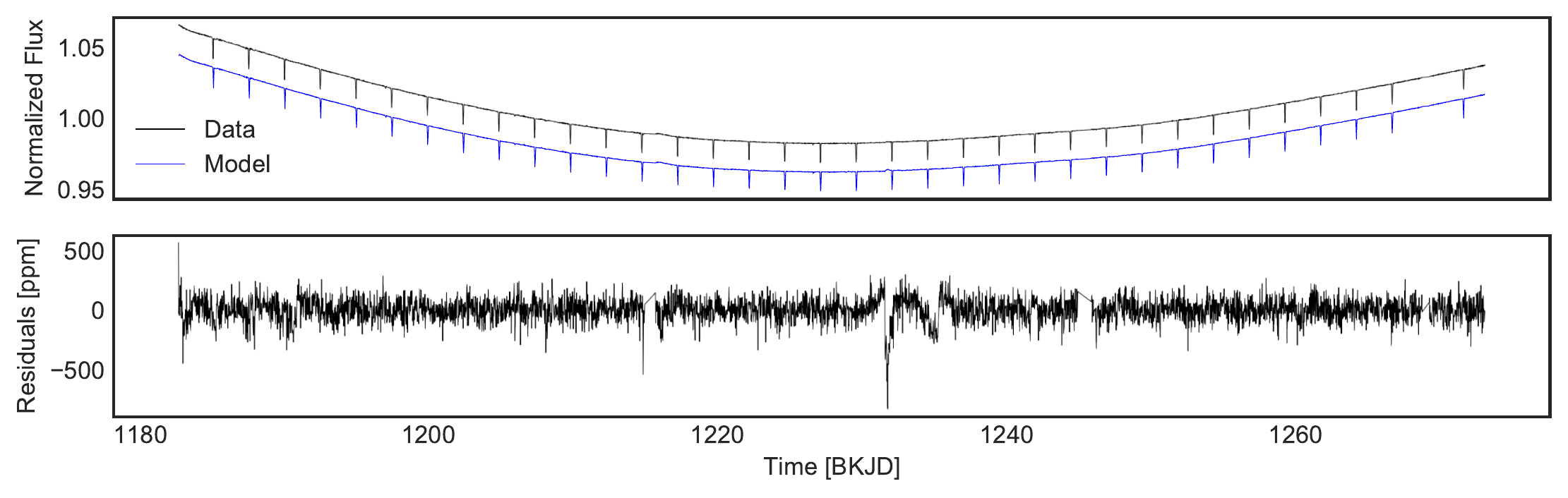}
    \caption{Example of the model fit described in Section~\ref{sec:model}. Shown is a single pixel of data from the Quarter 13 of transiting exoplanet Kepler-1b. (The model (blue) has been offset for clarity). Our model has excellent agreement with the data, as shown by the residuals in the lower panel. In this case we find our model typically has agreement at the $\sim$200$ppm$ level in each pixel. In order to build our full data-driven model, we produce models for every pixel in every quarter of \kepler data.}
    \label{fig:kepler1bdemo}
\end{figure*}

To demonstrate the detrending power of the model we have described, we apply it to the dataset on transiting exoplanet Kepler-1b (a.k.a TrES-2b) \citep{kepler1b}. Kepler-1b is an ideal test case for this model; the deep and short period transits, as well as the bright host star, make the astrophysical signal in each pixel strong. In addition, a transiting planet is expected to cause no wavelength dependent variability in the target flux. Even in the case of the hot planet of Kepler-1b, any emission from the planet would peak well outside the \kepler bandpass in the infra-red, and changes in limb darkening are anticipated to be small. Figure ~\ref{fig:kepler1bdemo} shows the model fit to a single pixel of Quarter 13 data of Kepler 1b. Our model has an exceptional fit to the data, providing an accurate fit at better than 200ppm in most cadences. Our model is similarly accurate in each pixel in the dataset.

\begin{figure*}
    \centering
    \includegraphics[width=\textwidth]{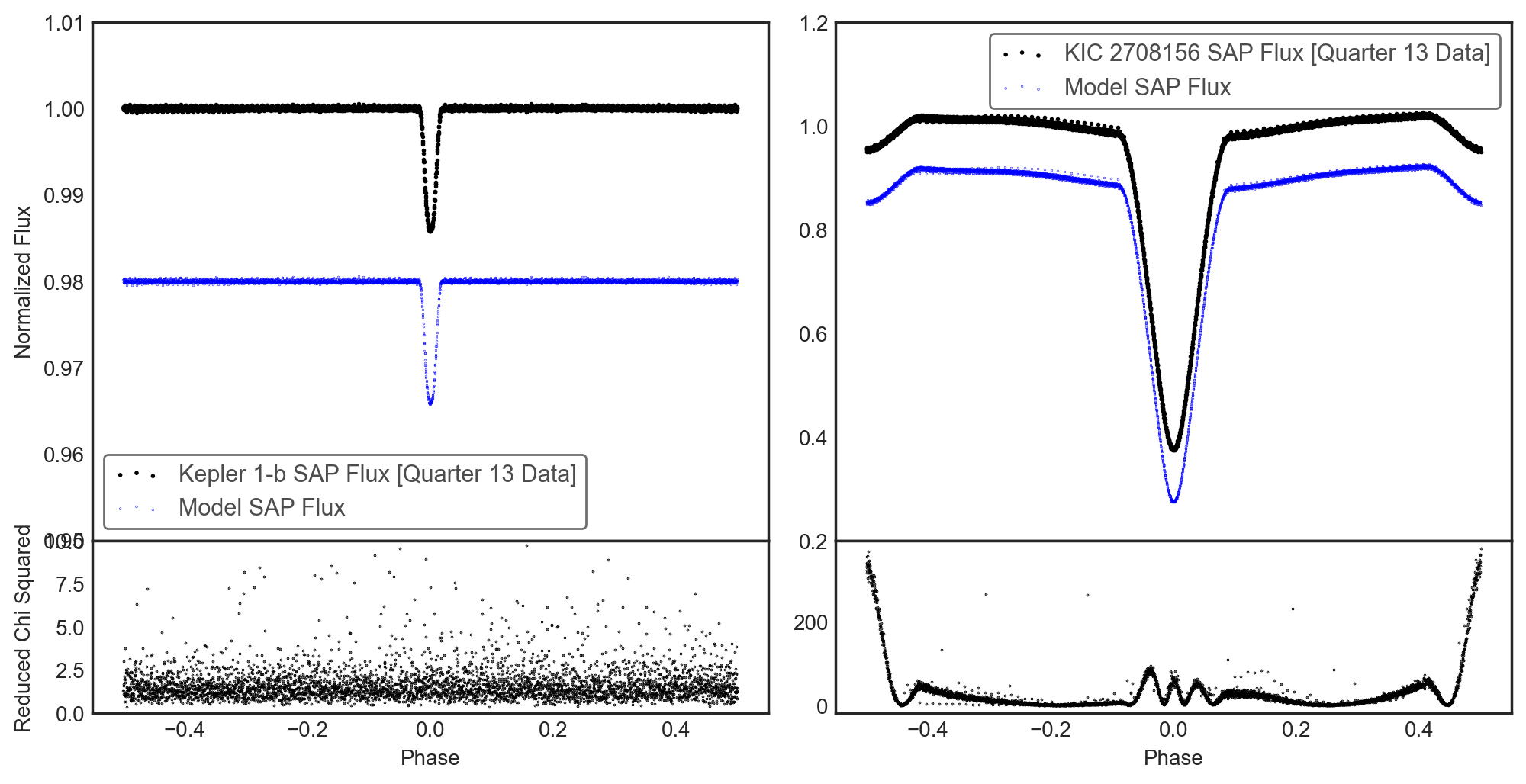}
    \caption{Example of our model fits to two different targets; Kepler-1b, an exoplanet with no expected wavelength dependent variability, and \target, an eclipsing binary where wavelength dependent variability is expected. Each target is folded at their respective orbital periods. The \kepler Simple Aperture Photometry (SAP) flux is shown in black in the upper panel for each target. The blue points show our model (offset for clarity), which is also summed over the same pixels in the aperture. The reduced chi squared fit of the model to the data is shown in the lower panel, summed over each pixel in the aperture. This is also folded at the orbital period of each target. Our model shows very good agreement with the time-series from the transiting planet target Kepler-1b, however the same model shows significantly poorer fits during the eclipses of the eclipsing binary target \target. This is evidence that the \kepler PSF shape is changing significantly during these eclipse events.}
    \label{fig:color_full}
\end{figure*}

The upper left panel of Figure~\ref{fig:color_full} shows the SAP flux for Kepler-1b during Quarter 13 of \kepler, folded at the orbital period of Kepler-1b. Our pixel level model is also summed over every pixel to create a model SAP flux light curve (blue). In the lower panel, the reduced chi squared statistic is shown for each cadence, showing that our model has good agreement at all phases of the light curve (reduced chi squared value of $\sim$ 1). There are no significant excesses or deviations from the model during transit, and crucially there is no correlation between the residuals and the transit signal. A single quarter has been chosen for demonstration, but our model is similarly effective in all quarters.

\subsection{Model example: \target}
\label{sec:exampleEB}

\begin{figure}
    \centering
    \includegraphics[width=0.5\textwidth]{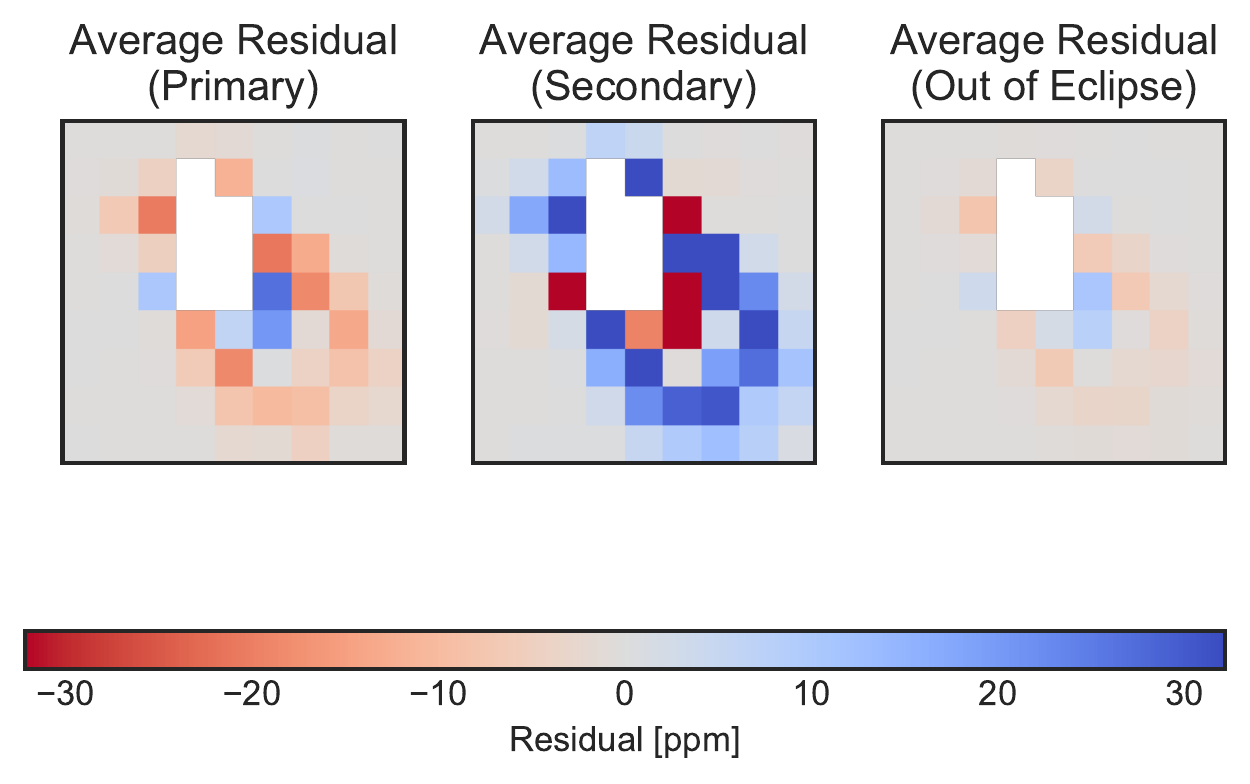}
    \caption{Example residuals of the model for \target. The median residuals during primary eclipse, secondary eclipse and out of eclipse are shown per pixel. Residuals are shown normalized to the average total flux of the target (measured through Simple Aperture Photometry). Here a red pixel indicates that the model systematically overestimates the flux, and blue indicates the model systematically underestimates the flux. We note that the residuals are strongest during eclipses. Pixels that were saturated have been omitted.}
    \label{fig:residualdemo}
\end{figure}

\begin{figure*}
    \centering
    \includegraphics[width=0.99\textwidth]{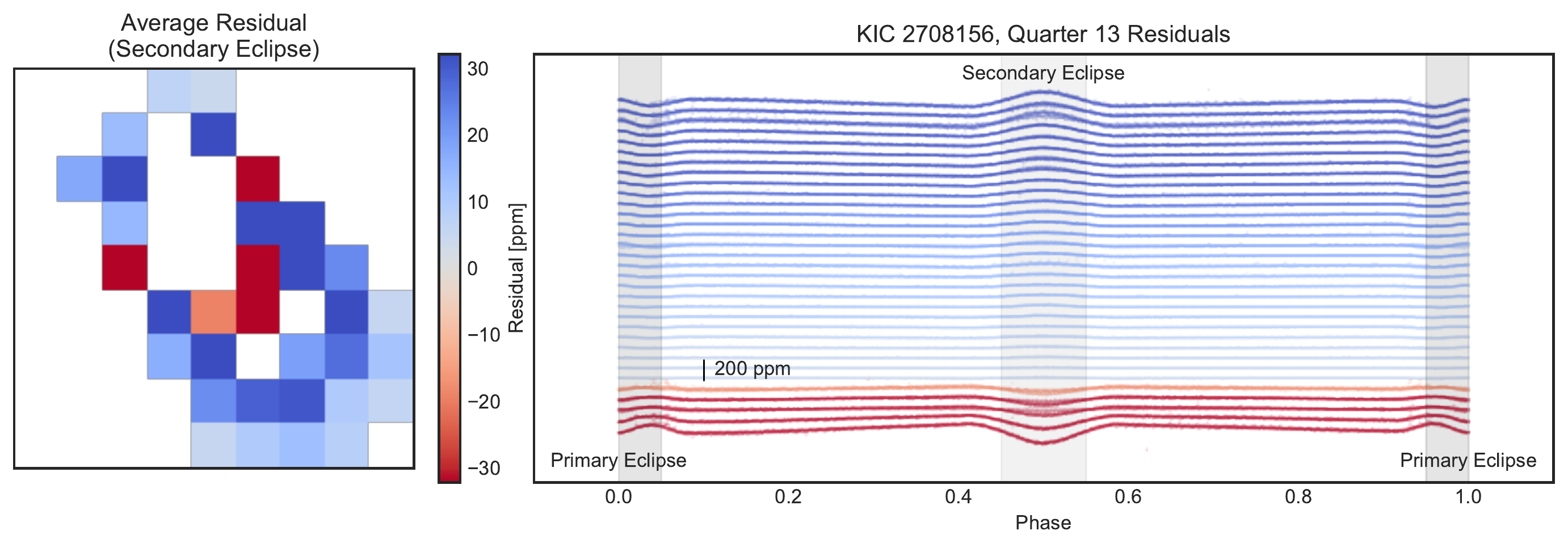}
    \caption{Example residuals of the model for \target, as a function of phase. \emph{Left:} The residual in parts per million averaged over all secondary eclipses during quarter 13, (middle panel from Figure ~\ref{fig:residualdemo}). Pixels are only shown where a change in secondary eclipse depth was identified at a $>3\sigma$ level. \emph{Right:} The residuals for each pixel, as a function of the phase of the eclipsing binary. Time series are color coded to indicate which pixel they originate from, and are ordered by secondary eclipse depth (with an offset for clarity). Secondary and primary eclipses have been indicated. Residuals are small ($\lesssim200ppm$) but correlated with the phase of the binary. The phase folded light curve for \target is shown in the right hand panel of Figure~\ref{fig:color_full}.}
    \label{fig:explainer}
\end{figure*}

The Kepler-1b dataset is not expected to have any wavelength dependence, since exoplanets are not bright enough to have significant emission in the \kepler bandpass. To find targets where wavelength dependence is both expected and predictable, we can look to eclipsing binaries. Eclipsing binaries are highly variable, and their astrophysical signal can easily be identified using only a single pixel of the broad PRF. Eclipsing binaries also have a predictable color change; if the two stars are different temperatures, they will have different apparent colors in the \kepler bandpass. As one star eclipses the other, the color of the target will change, as each star's contribution is downweighted when it is eclipsed. \target is an eclipsing binary, which was observed by \kepler during the prime mission, (parameters available in Table~\ref{tab:parameters}). The right panel of Figure~\ref{fig:color_full} shows our model fit to Quarter 13 data on \target, and shows that our model has a significantly poorer fit during the eclipses of this target. The true SAP flux is shown for \target, as well as the modeled SAP flux. The reduced chi squared of the model fit is shown in the lower panel, summed up over all pixels in the aperture. The reduced chi squared shows poor agreement between the model and the data, and crucially shows poor agreement at certain phases of the eclipsing binary. The model fits much more poorly during secondary eclipse, and primary eclipse.

\begin{figure*}
    \centering
    \includegraphics[width=0.8\textwidth]{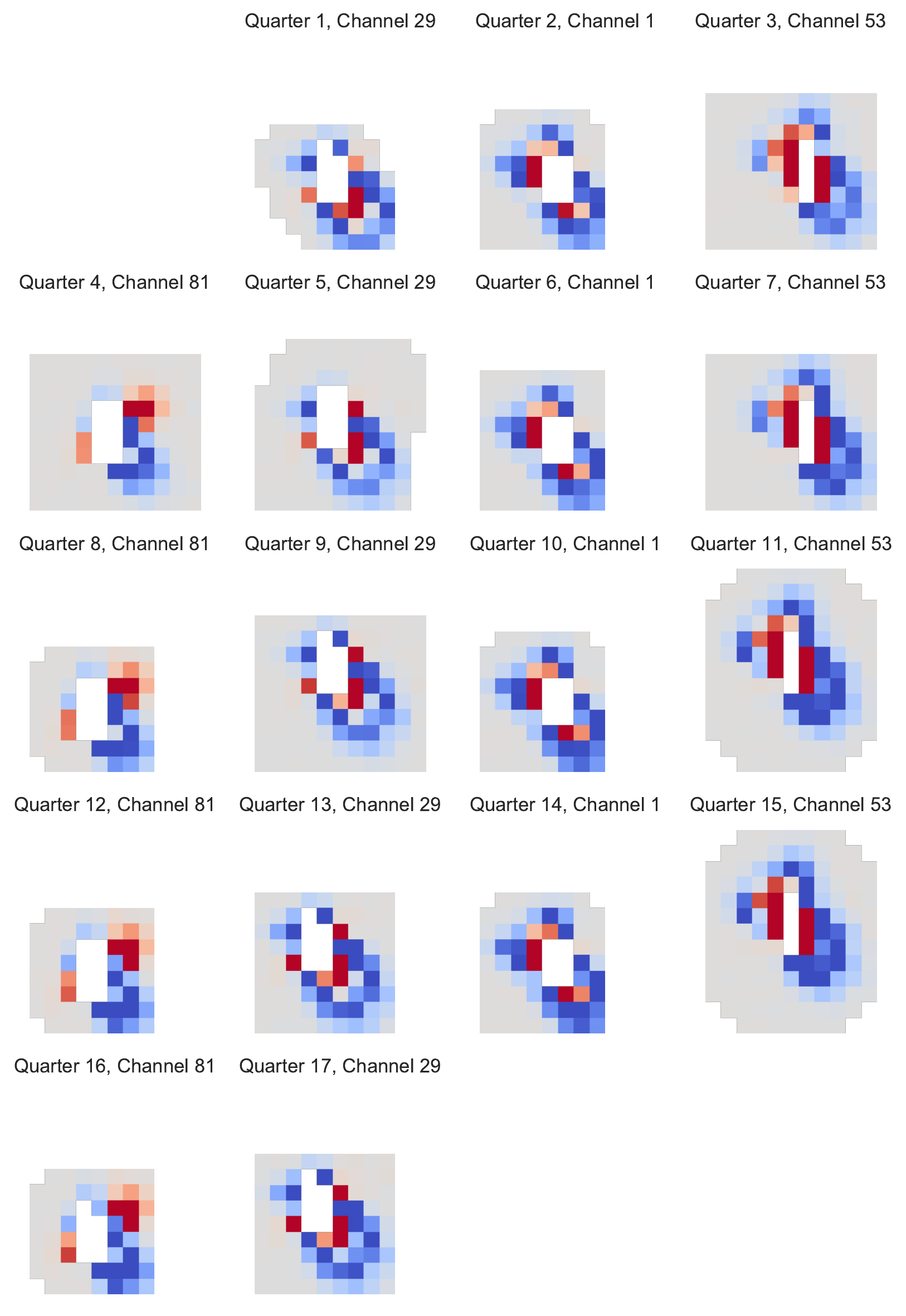}
    \caption{Example of residuals during secondary eclipse for every \kepler quarter of \target. Because of the observing strategy of the \kepler mission, the target is on a different channel (with a different PRF) each quarter of the year, but returns to the same channel every 4th quarter. As shown here, the residual PRF shape is different on each channel, as expected, but is almost identical during quarters that land on the same channel. Note: the color scaling has been fixed for clarity to show the same level of variability across all channels. The color map has been chosen to reflect the true color change; i.e. blue pixels indicate that the secondary eclipse depth residual became positive, indicating a bluer pixel light curve, red pixels indicate that the secondary eclipse depth residual became negative, indicating a redder pixel light curve. The wavelength calibration of these pixels in terms of wavelength is discussed in Section \ref{sec:calibrate}.}
    \label{fig:quarter_demo}
\end{figure*}

To show these residuals more clearly, Figure \ref{fig:residualdemo} shows the average residual for each pixel for \target during 1) primary eclipse 2) secondary eclipse 3) out of eclipse in Quarter 13. These images can be thought of as the residual between our PRF model and the data in pixel space. Given that we expect \target to have a color dependence, and that the PRF should be wavelength dependent, we would expect that the PRF shape will change during primary and secondary eclipse.

In Figure \ref{fig:residualdemo}, we see that 1) the PRF is, on average, broader during secondary eclipse 2) the PRF is, on average, narrower during primary eclipse and 3) the PRF shapes during primary and secondary eclipse are the inverse of each other and 4) the shape change during secondary eclipse is qualitatively similar to the theoretical PSF shape shown in Figure~\ref{fig:psf_demo}. Together, these facts imply that we are seeing PSF shape change due to color variability. Outside of eclipse we see that there is still a slight residual, and in Figure~\ref{fig:residualdemo} we see this is due to there being slight variations in the eclipsing binary phase curve in each pixel, implying that there may be color variability in the eclipsing binary phase curve.

Figure ~\ref{fig:explainer} shows the residuals, per pixel, as a function of phase in the eclipsing binary. Each pixel where a significant residual was detected during secondary eclipse is shown. While the residuals are small ($\lesssim200ppm$) they are evidently correlated with the eclipsing binary phase. We see there is a clear change in eclipse depth, and that the ingress and egress of the primary eclipse is changing significantly in each pixel. 

\cite{ebliterature} provide temperature estimates for a small sample of bright eclipsing binaries. According to \cite{ebliterature}, the primary of the \target system has an effective temperature of 11061$K$, and the secondary has effective temperature of 5671$K$. These two different temperatures will result in each eclipse depth (both primary and secondary) having a strong color dependence. In the case of \target, we expect the target will become redder during primary eclipse, as the hotter star is eclipsed by the cooler star. Similarly, we expect the target will become bluer during secondary eclipse. Given that the target will cycle between redder and bluer phases, it is compelling that the primary and secondary PRF residuals shown in Figures ~\ref{fig:residualdemo} and ~\ref{fig:explainer} are the inverse of each other.


To further demonstrate the PRF shape, we show the residuals in each \kepler quarter. The \kepler spacecraft rotated 90 degrees every quarter, and so each target falls on a different channel of the detector every season. Every fourth quarter, the target will fall on the same channel, in almost the same place. However, because of temperature changes and velocity aberration, the target center may have moved $\lesssim$ 1 pixel. Because \target falls on a different channel each season, we would expect the PRF to be different each season (owing to the unique PSF shape and detector systematics of each channel). However, during quarters that fall on the same channel, we would expect the PRF, and thus the residual PRF, to be largely the same. Figure \ref{fig:quarter_demo} shows the residual PRF during secondary eclipse (the central panel from Figure~\ref{fig:residualdemo}) for every quarter of \kepler data on \target. The color scaling is fixed between each quarter to show the same level of variability. We see that all quarters that fall on the same channel have extremely similar residual PRF shapes during secondary eclipse, further supporting that the effect is real, and intrinsic to the target PRF. The color map has been chosen in this figure to reflect the true color change; i.e. blue pixels indicate that the secondary eclipse depth residual became positive, indicating a bluer pixel light curve, red pixels indicate that the secondary eclipse depth residual became negative, indicating a redder pixel light curve. Figure~\ref{fig:residualdemo} shows that pixels invert residual during primary and secondary, as the eclipsing binary changes color. We also see the PRF shape is also qualitatively similar to Figure~\ref{fig:psf_demo}; there is a small central cluster of pixels (red) which are inverted compared to the broader halo of pixels (blue).


\subsection{The "brighter-fatter" effect}

In the above sections, we have shown the PRF shape for a single eclipsing binary is highly correlated with the phase of the eclipsing binary. There exists a known systematic effect that may cause the PRF shape to change in a way that is correlated with astrophysical signal, known as the "brighter-fatter" effect. This systematic effect is well documented \citep[e.g.][]{bf0, bf1, bf2}, and causes the shape of PSFs to get broader as sources become brighter, due to increased electrostatic repulsion as a pixel accumulates charge.

In the above sections, we have purposefully chosen a bright eclipsing binary with large eclipse depth. We have included terms to account for non-linearity in each pixel, which should in theory entirely account for the brighter-fatter effect (as we are treating each pixel independently). However, the flux of the target is changing significantly, and we must verify that the shape change we have identified is not due to the brighter-fatter effect. To do this, we show the reduced chi-squared of the model for \target as a function of flux in Figure~\ref{fig:non-linear}. (also shown in Figure~\ref{fig:color_full} as a function of time).  

In Figure~\ref{fig:non-linear}, the primary and secondary eclipses have been highlighted in red and blue respectively. In reduced chi squared, the primary and secondary eclipses clearly follow two tracks, with secondary eclipse having a much higher reduced chi squared than the primary eclipse at a flux of ~0.9. Crucially, there are several instances where the flux of the primary and secondary eclipse are the same (normalized flux $\approx0.85 - 1$), but the reduced chi squared is significantly higher for the secondary eclipse. Despite having the same measured flux on the detector, the model residual is more significant during secondary eclipse; the pixel-level model fits more poorly during secondary eclipse, indicating a more significant shape change. Given that the measured flux is the same, the brighter-fatter effect is not able to account for this difference in PRF shape between primary and secondary eclipse, and we can discount this as the source of the PRF shape change.

\section{Verification of Color Dependence Using Eclipsing Binaries}
\label{sec:EBs}

Based on Section~\ref{sec:exampleEB}, there is evidence that \kepler data of \target contains wavelength-dependent variability. While Section~\ref{sec:exampleEB} contains compelling evidence that we are detecting photons at different wavelengths in each pixel, there could be further instrument systematics that we are unaware of that are causing this effect in this specific target. To verify that the effect is due to a wavelength-dependent PRF, we can use a sample of bright eclipsing binaries where we have temperature estimates for each star in the binary. Each binary has its own set of temperatures, and unique orbital parameters, and as such will have a unique color variability as a function of phase. 

In this section we will demonstrate the wavelength dependence of the \kepler PRF by \begin{enumerate}
\item \textbf{Modeling the systematics for each target, for every pixel and every quarter.} We will build the same model as discussed in Section~\ref{sec:keplermodel}, modeling the astrophysics and variable instrument systematics in every pixel light curve. 
\item \textbf{Building theoretical time series for each eclipsing binary, evaluated at a coarse wavelength grid.} These models will include the theoretical, expected change in eclipse depths due to the change in luminosity of each star, evaluated at a grid of wavelengths. To build these eclipsing binary time-series models we will assume simple blackbody emission from each star in the eclipsing binary. This eclipsing binary time-series model is discussed in Section~\ref{sec:binarymodel}. 
\item \textbf{Compare the measured eclipse depths in each residual pixel time series with the anticipated theoretical eclipsing binary light curves at each wavelength.} We will show the residuals of the data and our pixel level model in each pixel contain significant changes in eclipse depth (both secondary and primary), and that these residuals correlated with the anticipated wavelength variability predicted by a basic eclipsing binary time-series model. This comparison is discussed in Section~\ref{sec:resultebs}.
\end{enumerate}

\subsection{Sample Selection}
In this work we use a small literature sample of eclipsing binaries with temperature estimates for each component from \cite{ebliterature}. From that sample, we exclude systems with non-zero eccentricities, obvious eclipse timing variations, and uncertain temperatures. Our reduced sample of 27 eclipsing binaries is given in Table~\ref{tab:parameters} in the Appendix. All targets in this sample are bright \kepler eclipsing binaries at short periods. We select this small sample only as a proof of concept for the wavelength dependent PRF, but in theory this effect would apply to all bright eclipsing binaries observed with \kepler.

\subsection{Eclipsing Binary Modeling}
\label{sec:binarymodel}
For each of the eclipsing binaries in Table~\ref{tab:parameters} we fit the Simple Aperture Photometry light curve to find the best fit orbital parameters (including radii and impact parameter). We use the analytic light curve modeling code \texttt{starry}\footnote{https://rodluger.github.io/starry/v1.0.0/} \citep{starry} to find the best fit solution. In order to model phase curve variation, we use simple models from \cite{beer}, given as

\begin{equation}
\label{eq:phase}
a_{re}\cos(2\pi\phi + \pi) + a_{el}\cos(4\pi\phi + \pi)  + a_{do}\sin(2\pi\phi + \pi) + c
\end{equation}
where $\phi$ is the phase of the eclipsing binary, $a_{re}$, $a_{el}$ and $a_{do}$ are free parameters for the weights of the reflected, ellipsoidal and Doppler beaming contributions respectively, and $c$ is a constant.
Using literature values of effective temperatures from \cite{ebliterature}, and assuming a blackbody emission for each component, we then predict the eclipsing binary time series at 400$nm$, 500$nm$, 700$nm$ and 800$nm$.  Only the luminosity of each star is varied at each wavelength point, and any change in the radius of the star is assumed to be negligible.



\subsection{Identifying a trend across the eclipsing binary catalog}
\label{sec:resultebs}

Using this grid of wavelength dependent light curves, we can now test whether the \kepler data of the eclipsing binary sample shows consistent evidence of color variability, in the same way as \target. To test this, we first build pixel model light curves for every target, for every pixel and quarter, using the same methods as discussed in Section~\ref{sec:keplermodel}. We then build the residuals of the true \kepler data and our data-driven pixel time-series model. 

For targets where there is a strong wavelength dependence, we expect the secondary eclipse depth residual to be large. But, since each pixel may have a different effective wavelength, the residual eclipse depth varies between pixels (see the right panel of Figure ~\ref{fig:explainer}). We define the ``\textbf{measured eclipse depth variation}'' in this dataset as the standard deviation of the residual eclipse depths, measured in each pixel for a single quarter observation, in pixels where a residual is detected at $\ge$3$\sigma$ significance. This is the standard deviation of the values in the left panel of Figure ~\ref{fig:explainer}. We take this measurement once per quarter. In this section we choose the standard deviation as our metric for the change in eclipse depth, which allows us to combine the residuals in all pixels into a single, simple metric. (In Section~\ref{sec:coloraperturephotometry} we use a more sophisticated approach to extract wavelength information.)

Each eclipsing binary has a unique expected color variability; binaries containing stars with similar temperature will have a lower magnitude wavelength variability than stars with extreme temperatures, and binaries with deeper eclipses will have a larger relative change. To understand our sample, we need a single numeric value to assess the magnitude of the expected, theoretical wavelength dependence. We define the \textbf{theoretical eclipse depth variation} as the difference between the eclipsing binary time-series model at 500$nm$ and 700$nm$, averaged during secondary eclipse. We choose wavelengths that are slightly bluer and slightly redder than the \kepler bandpass center. This provides a single, quantitative estimate of the magnitude of color variability anticipated for each target, based on the temperatures and orbital parameters of the star components. These wavelength values have been chosen somewhat arbitrarily to construct this simple numeric estimator, in order to demonstrate the consistent wavelength dependence in our eclipsing binary sample.

\begin{figure}
    \centering
    \includegraphics[width=0.5\textwidth]{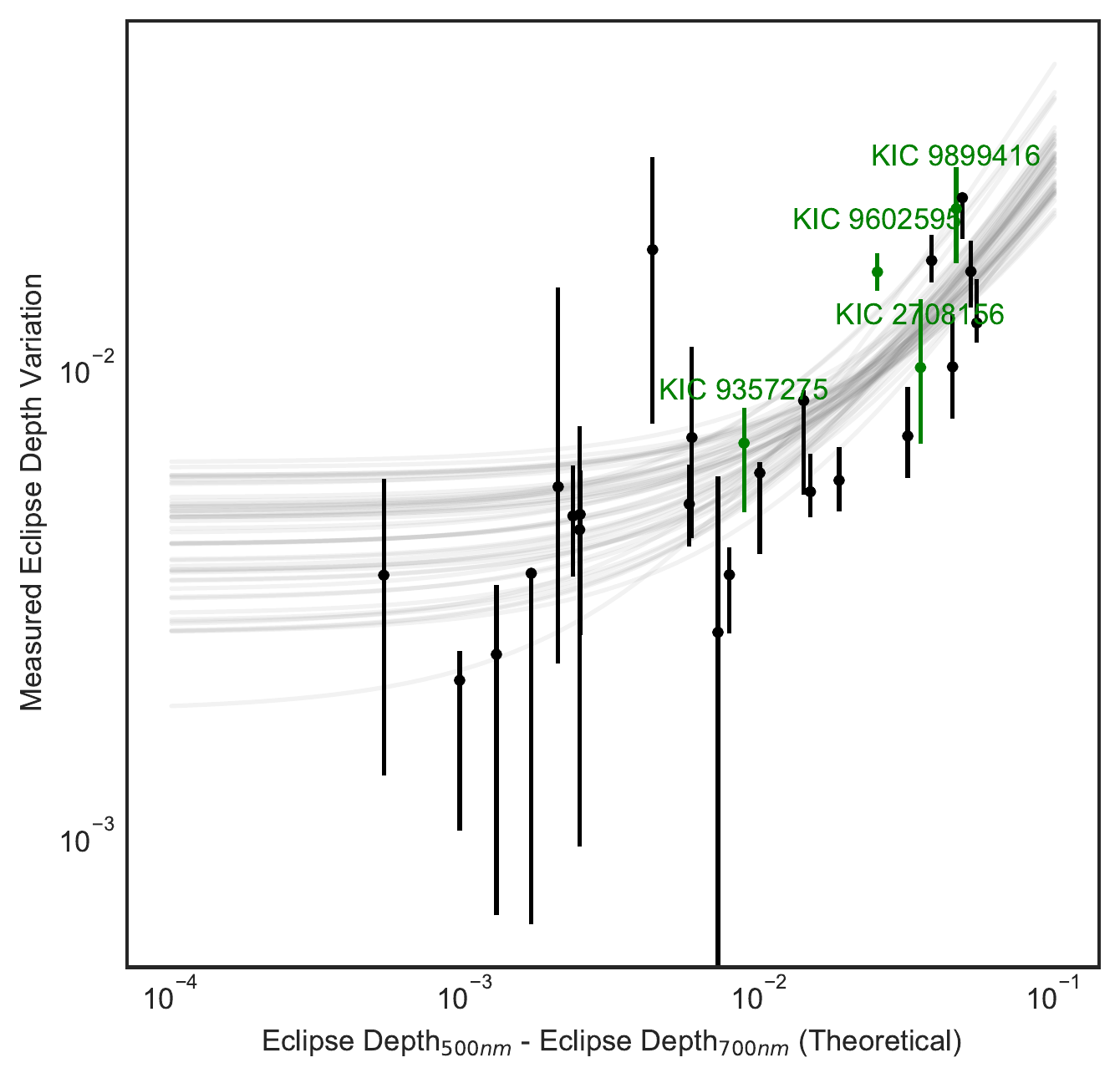}
    \caption{Theoretical eclipse depth variation based on eclipsing binary models for our sample evaluated at 500$nm$ and 700$nm$, compared with the measured eclipse depth variation in the residuals between the data and our pixel-level systematics model. The mean value for each target is plotted with uncertainties, which are given by the standard deviation measured over each quarter. 50 realizations of the best fit linear models are shown in grey. There is a clear correlation between the theoretical expected variation due to wavelength dependence and the measured residuals, establishing that the \kepler data is weakly wavelength dependent. Targets highlighted in green are plotted in Figure~\ref{fig:wavelength_demos}.}
    \label{fig:correlation}
\end{figure}

Figure~\ref{fig:correlation} shows the correlation between the \textbf{theoretical eclipse depth variation} and the \textbf{measured eclipse depth variation}, for each target. We plot the average measured eclipse depth variation across all \kepler quarters, with uncertainties given by the standard deviation of the measurement across all quarters. As shown in Figure~\ref{fig:correlation}, we find a statistically significant correlation between the measured variation in eclipse depth, and the predicted theoretical eclipse depth variation across our sample. Using a Pearson Correlation test we find there is a ($\sim 3\sigma$) correlation between our theoretical prediction of the color variability of our sample, and the measured PRF shape change. This demonstrates that the wavelength dependent effects shown in Section~\ref{sec:exampleEB} are not simply coincidence, but true evidence of the wavelength dependence of \kepler observations.

\begin{figure*}
    \centering
    \includegraphics[width=0.95\textwidth]{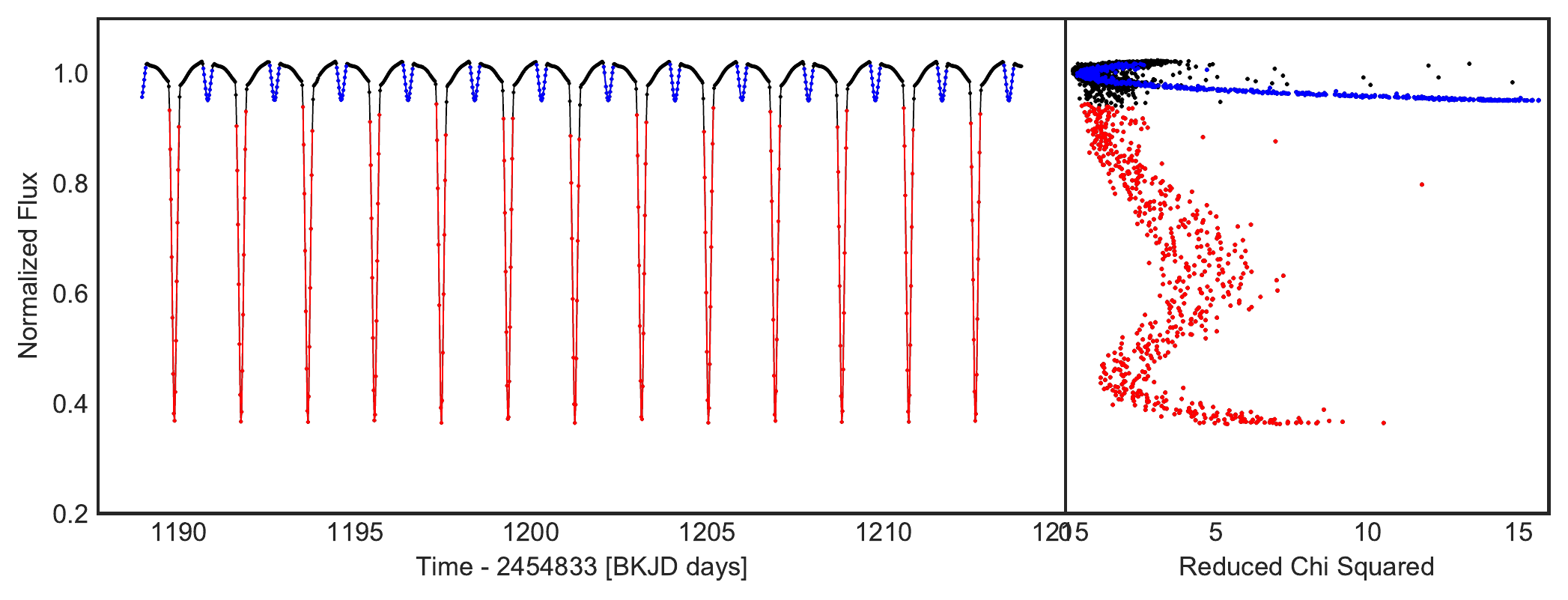}
    \caption{\emph{Left:} The SAP flux for \target as a function of time. \emph{Right:} The SAP flux for \target as a function of reduced chi squared. The primary eclipses have been highlighted in red, and the secondary eclipses have been highlighted in blue. The primary and secondary eclipses clearly follow two tracks in reduced chi squared, with the secondary eclipse having generally higher reduced chi squared as a function of flux. Given that there are several instances where the flux from primary eclipse and secondary eclipse is measured to be the same (normalized flux $\approx 0.85-1$), we understand the pixel-level model fit to be "worse" during secondary eclipse. As such, the PRF shape change we identify in this work can not be attributed to simple systematic effects due to non-linearity of flux, including the brighter-fatter effect, which would be agnostic of the "phase" of the binary.}
    \label{fig:non-linear}
\end{figure*}

\subsection{Coarse Wavelength Calibration}
\label{sec:calibrate}

It is beyond the scope of this initial work to fully model the PRF shape of \kepler. Fully modeling the PRF would require an investigation in the entire dataset, and would provide a PRF model as a function of magnitude, spatial position on the detector and wavelength. However, using our sample, we can provide a coarse wavelength calibration, and estimate the range of effective wavelengths for each pixel in our sample. 

\begin{figure}
    \centering
    \includegraphics[width=0.5\textwidth]{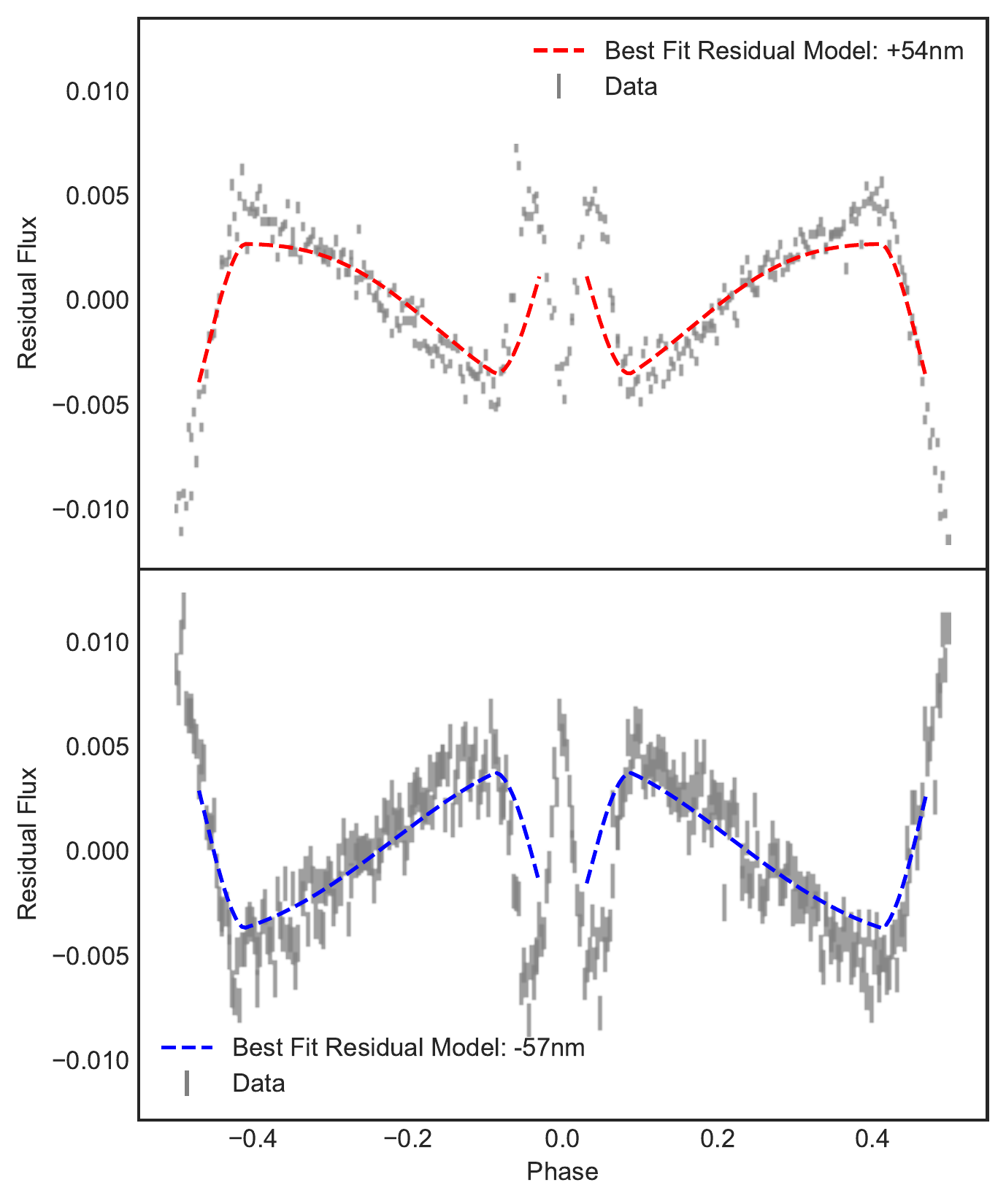}
    \caption{Two example residuals in pixels from Quarter 1 data of \target. Grey points show residuals between the \kepler data and our PRF model, folded at the orbital period of \target. Two pixels have been chosen which show a deeper and shallower secondary eclipse compared with the simple aperture photometry flux. Fitting our eclipsing binary light curve model to the residuals we find that these pixels are recording flux at $\pm \approx$ 50$nm$. Note that there is a significant deviation from the model at phase of 0, which we attribute to a change in the limb darkening. The model is not fit at phases close to 0 or $\pm$0.5, in order to avoid fitting this limb darkening variation. In future analyses, it would be possible to build a model that takes this wavelength variation into account.}
    \label{fig:examplepixelfit}
\end{figure}

To provide a wavelength calibration we will fit the residuals of the \kepler data less our PRF model (Section~\ref{sec:model}) with our wavelength dependent eclipsing binary time-series model, for every pixel, and every quarter. We use a grid of pre-computed eclipsing binary models for each target, evaluated at 300-900$nm$ with a grid spacing of 100$nm$. We then interpolate between these grid points to find the best fitting wavelength model. We additionally fit phase curve variations using Equation~\ref{eq:phase}, to allow for wavelength dependent phase curve variation (we consider this to be a nuisance parameter in this work). Using this model we estimate the effective wavelength of each pixel. An example of this fit is shown in Figure~\ref{fig:examplepixelfit} for two example pixels of \target. In Figure~\ref{fig:examplepixelfit} it is clear at phase of 0 there is a significant, sharp change in the eclipse depth. We attribute this to wavelength dependent changes in limb darkening. (This effect is more clearly shown in Figure ~\ref{fig:wavelength_demos}). Our model here is approximate, and could be improved by including physically motivated wavelength dependent limb darkening and phase curve variation parameters. We therefore mask out this part of the phase curve when fitting out eclipsing binary model.

\begin{figure}
    \centering
    \includegraphics[width=0.5\textwidth]{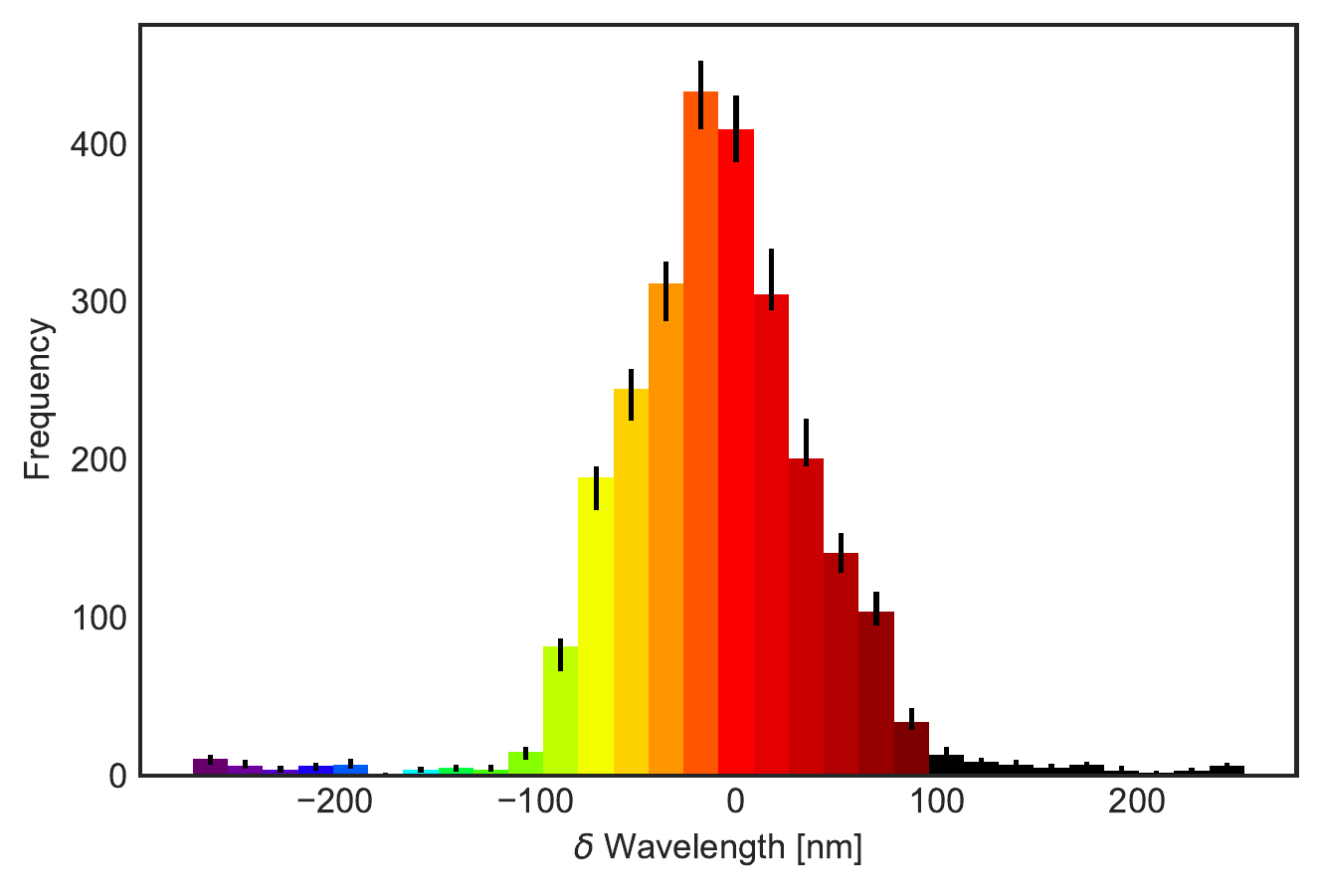}
    \caption{Histogram of all estimated wavelengths in the dataset. We estimate the wavelength for each pixel and \kepler quarter of each EB in our sample, using the procedure outlined in Section~\ref{sec:calibrate}. For illustrative purposes, visual colors corresponding to each wavelength are plotted, (assuming the \kepler bandpass center is at 646.1$nm$). We find a standard deviation across the sample of $\approx$40$nm$.}
    \label{fig:rainbow}
\end{figure}

Figure \ref{fig:rainbow} shows a histogram of best fit wavelengths to every pixel in our sample, (i.e. every pixel, in every quarter, for every target). Using a simple Chi-squared test, we remove pixels where the eclipsing binary model is a poorer fit than a flat line. A value of 0 indicates that there is no significant difference between the SAP flux light curve and the pixel light curve. We find the distribution of wavelengths to be approximately Gaussian, with a standard deviation of 40nm. Figure~\ref{fig:rainbow} shows that, for bright eclipsing binaries observed by \kepler, a broad range of wavelengths are accessible. 

\section{Color Aperture Photometry}
\label{sec:coloraperturephotometry}
\begin{figure*}
    \centering
    \includegraphics[width=\textwidth]{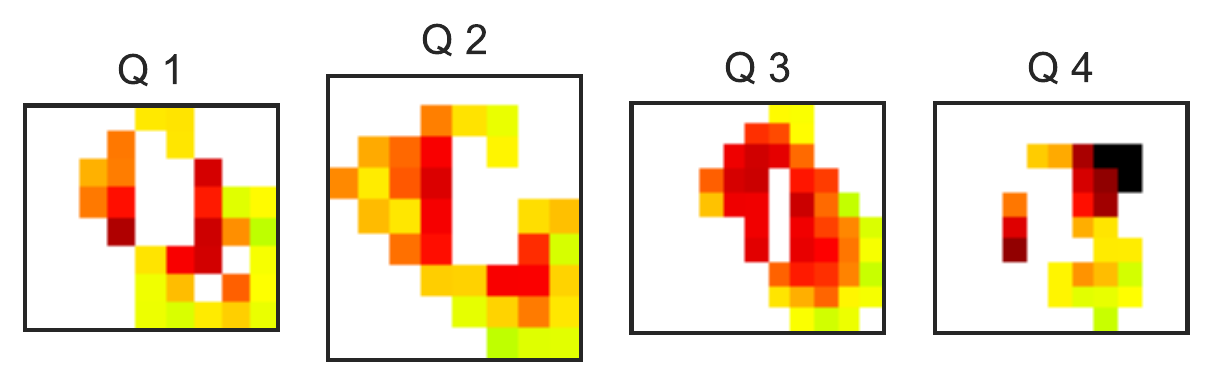}
    \caption{The "true" colors for each pixel in the first four quarters of \target, estimated using our eclipsing binary model. White pixels are where no fit could be achieved, either due to low signal to noise, or saturation. Black pixels are (slightly) redder than the human eye can detect. This distribution is qualitatively similar to the PSF model shown in Figure 2, the "bluer" pixels are at the edges, and "redder" pixels are more concentrated towards the middle of the PRF.}
    \label{fig:true_color}
\end{figure*}

As shown in the above sections, for eclipsing binaries we area able to create a coarse wavelength calibration, and identify pixels that are similar in their wavelength response. For illustrative purposes, we show the "true" color of each pixel for \target in Figure ~\ref{fig:true_color}. We are now able to group these pixels together and perform "Color Aperture Photometry" (CAP). To do this, we group pixels together and create multiple "apertures" over pixels with similar measured wavelengths, across all quarters. In the context of Figure~\ref{fig:true_color}, we would select pixels with "similar" colors (defined by some wavelength bin), to build a Color Aperture. The pixels in each color aperture are then summed to create multiple time-series, with one time-series per aperture (wavelength bin). We can then compare these time series to show the wavelength-resolved, photometric variability of each target.

\begin{figure*}
    \centering
    \includegraphics[width=\textwidth]{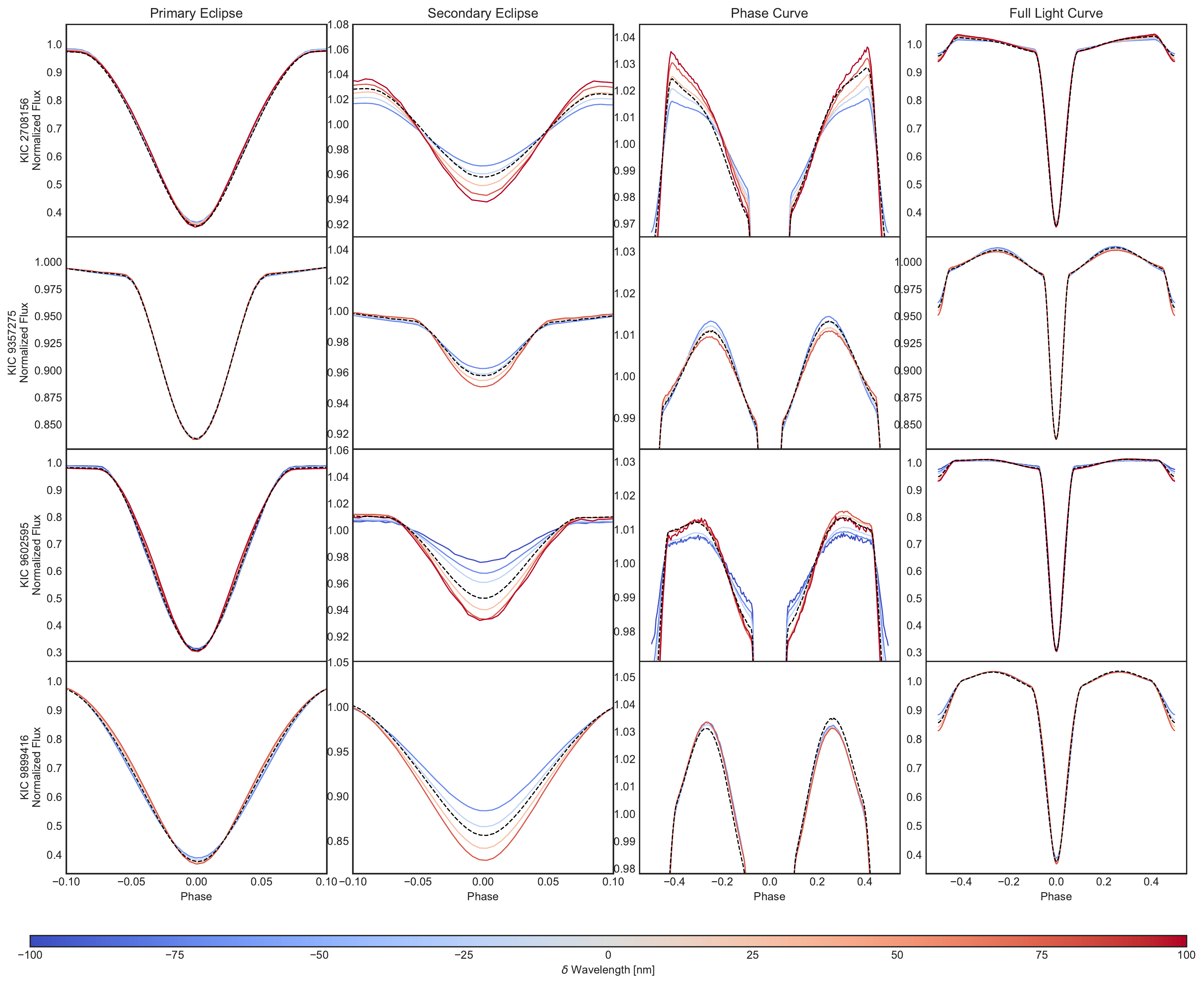}
    \caption{Four example eclipsing binaries from our sample with clear variations due to the chromatic \kepler PRF. Shown are "Color Aperture Photometry" (CAP) light curves of each target, for every quarter and pixel of \kepler data, binned into the approximate wavelength bin found by this work. Each of the panels in this plot have been x- and y-scaled to show the effects of chromaticity in clearly. \emph{First Column}: Rescaled plot to show primary eclipse variations. The subtle effects of changes in limb darkening are seen in each panel. \emph{Second Column}: Rescaled plot to show the secondary eclipse. Here the secondary eclipse depth is clearly varying across each wavelength, (this is the effect used to select the wavelength bins to build the CAP light curves.) It is also possible to see the subtle effects of variable limb darkening in these light curves. \emph{Third Column}: Rescaled plot to show the phase curve variation. In this plot it is clear that each binary has a unique phase curve, with a unique wavelength dependent response. \emph{Fourth Column}: Full eclipsing binary light curve, without rescaling. The color bar shows the wavelength of each light curve, relative to the \kepler bandpass center. The dashed line indicates the SAP flux light curve.}
    \label{fig:wavelength_demos}
\end{figure*}

Figure ~\ref{fig:wavelength_demos} shows CAP light curves of 4 eclipsing binaries from our sample. (These targets are also highlighted in Figure~\ref{fig:correlation}.) Pixels were grouped into apertures in 25$nm$ bins, and summed over all quarters. Light curves have been grouped into panels for clarity, showing the primary and secondary eclipses, phase curve variations, and full light curve. As shown in that figure, 1) the secondary eclipse depth varies significantly in each wavelength bin 2) the phase curve varies in each wavelength bin 3) limb darkening during both primary and secondary eclipse varies. Each of these three components are expected to have a wavelength dependence, and the fact that each of these components are independently changing is the final piece of evidence that the PRF of \kepler is wavelength dependent, and that it is possible to extract coarse, multi-wavelength photometry from \kepler data.

The CAP light curves in this work are presented only as examples, but further work could use the  techniques presented here to extract multi-wavelength photometry of the \kepler eclipsing binary sample, and estimate their fundamental properties (such as temperature ratio, or modeling ellipsoidal variations and doppler beaming etc). This work could be expanded to include eclipsing binaries beyond the small sample used in this work.

Eclipsing binaries present an ideal case for CAP; it is possible to use the data itself to build the correct pixel apertures for each wavelength. By using the methods above, we have identified which pixels contain "similar" secondary eclipse depths (well motivated by our understanding of eclipsing binary time-series), and summed those in individual apertures. However, many targets exist in the \kepler dataset where there is no predictable simple model for anticipated color variation, and we will not be able to correctly select the CAP apertures using the target data. To build CAP apertures in such cases, we would require an in-depth investigation into the PRF across a \kepler channel, in order to identify the best CAP apertures, agnostic of the target. However, as discussed in Section \ref{sec:uses}, there are other uses for the model presented in this work that do not require building CAP apertures.

\section{Discussion}
\label{sec:discussion}

In the Section~\ref{sec:EBs} we have verified, using a sample of eclipsing binaries, that there is a clear wavelength dependence in \kepler data. The evidence presented in this work can be summed up as

\begin{itemize}
	\item{A significant PRF shape change is identified in \kepler data of eclipsing binaries, which correlates with the eclipses and phase curve variations (Figure~\ref{fig:color_full}). The change in PRF shape can not be explained by simple instrument systematics (Figure~\ref{fig:non-linear}).}
	\item{The PRF shape change is qualitatively similar (tight central region, surrounded by a broad halo)} to the expected theoretical PRF shape based on models of the chromatic aberration from the KIH. (Figure ~\ref{fig:residualdemo})
	\item{The measured PRF shape change shows significant correlation with the theoretical, expected PRF shape change across and sample of 27 eclipsing binaries, despite each binary having unique orbital parameters, spectral types, and instrument systematics. (Figure ~\ref{fig:correlation})}
	\item{The secondary eclipse, primary eclipse, phase curve, and limb darkening of eclipsing binaries in our sample vary simultaneous, at different magnitudes within each eclipsing binary. (Figure ~\ref{fig:wavelength_demos})}
	\item{The variations in secondary eclipse depth and primary eclipse depth can be fit simultaneously across our entire sample with a well motivated eclipsing binary model, including fixed temperatures for each star component from literature, in order to derive consistent wavelengths for each pixel that are Gaussian distributed with a narrow wavelength range (standard deviation of 40$nm$). (Figure ~\ref{fig:rainbow})}
\end{itemize}

Given this evidence, we assert that the \kepler data are weakly chromatic. In this section we discuss some limitations and possible uses of this measurable chromaticity.

\subsection{Limitations}
\label{sec:limitations}

The model for the PRF we have described here is limited. In particular, we have ignored all correlations, both in time between cadences and between pixels. In reality, 1) the systematics in consecutive cadences are highly correlated, and 2) pixels that fall close to each other in the PRF are highly correlated. In addition, observations that fall on the same channel (and therefore similar pixels) of the \kepler detector should be correlated, as they experience similar systematics and exhibit similar PRF shapes. Secondly, instrument effects, such as undershoot, column-dependent smear and row-dependent bias could may cause significant correlation between pixels \citep[see][]{kih}, which we don't account for in this work for simplicity. By including all these sources of correlations, it may be possible to build a more predictive model, applicable to targets beyond bright eclipsing binaries. 

Our data-driven approach fits a model to each pixel, and each new target requires a unique model. As such, using our current model, it is not possible to approximate the effective wavelength of an arbitrary pixel on the detector. By building a fully calibrated PRF model based on all the \kepler data it is theoretically possible to estimate the effective wavelength of each pixel. We leave this as further work. Using the work presented here, and a large enough sample of eclipsing binaries, it may be possible to create a model for CAP apertures, and use those apertures to build coarse color photometry from \kepler.

This work has shown that there is a color dependence in the \kepler PRF, and each pixel has a different effective wavelength. We have not explored the limits of this effect in terms of magnitude, and have restricted our study to only bright targets with extreme and predictable color variation. Further work is required to study the chromaticity of the \kepler PRF, and find the magnitude limit of the effect across the \kepler sample.

Finally, the \kepler bandpass is in fact a wide, spanning 400$nm$ (see KIH). While we have identified the approximate effective wavelength of pixels in our sample, this is an oversimplification. In fact, each pixel should be thought of as having a unique bandpass, which is the compound effect of all of the optical systems in the telescope, and which is different to that of the average \kepler bandpass. Further investigation is needed to understand the differences between our simplistic approach, and a thorough analysis that takes into account the possibility of a variable bandpass in each pixel.

\subsection{Example Use Cases}
\label{sec:uses}

Despite the considerations discussed in Section~\ref{sec:limitations}, and despite the limited wavelength range of the chromaticity effect, there are several ways that the community can capitalize on this coarse multi-wavelength photometry. In this section we discuss some of the ways that this effect could be employed to find new insight using the archival \kepler data, including gaining a better understanding of eclipsing binaries across multiple wavelengths, vetting exoplanet candidates, identifying and characterizing flares, and studying pulsators. There are numerous further uses for this coarse wavelength photometry that go beyond what is discussed here, including better understanding reddening in young stars, the color dependence of supernovae, or wavelength dependent asteroseismology measurements. This coarse multi-wavelength photometry, in conjunction with \kepler's extreme precision, can be used both for inference and for classification of objects. For example, by encorporating pixel-level modeling into machine-learning classification algorithms (e.g. \cite{vanderburg-and-shallue}) it may be possible to more accurately classify true transit events. 

\subsubsection{Classification: Exoplanet Vetting}
\begin{figure*}
    \centering
    \includegraphics[width=0.95\textwidth]{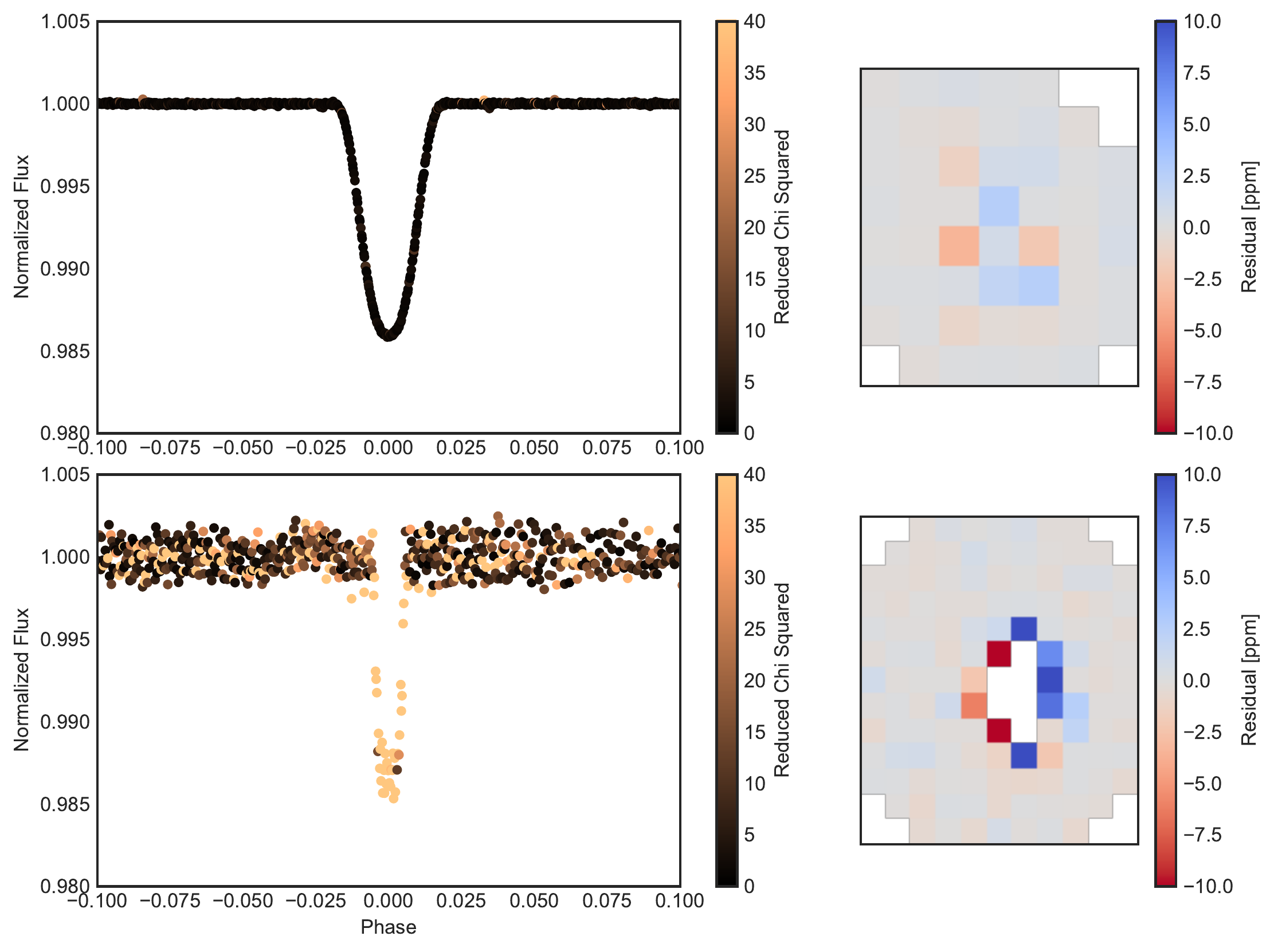}
    \caption{Demonstration of how the model presented in this work can be used to identify False Positive planet candidates in \kepler data. \emph{Left:} Folded light curve for a single quarter of \kepler data on Kepler-1b (top) and eclipsing binary KIC 8703887 (bottom). KIC 8703887 has been chosen for its similar eclipse depth and duration to exoplanet Kepler-1b, as well as similar apparent magnitude (Kepler-1b has a \kepler magnitude of 11.34 and KIC 8703887 has a \kepler magnitude of 11.05). The color bar in these panels shows the reduced chi squared goodness of fit metric between the data and our pixel level model. The eclipsing binary KIC 8703887 could be confused with a transiting planet using the time-series alone. However, the reduce chi squared shows a significant deviation from the model during transit, indicating a PRF shape change.  The true planet does not show a significant shape change. We attribute this to the color variation in the eclipsing binary light curve. \emph{Right:} Residual between the data and our pixel level model during transit, showing the significant deviation for the eclipsing binary and good agreement for the true planet. Using the reduced chi squared metric, we can easily identify which of these targets is the true planet, and which is a grazing eclipsing binary.}
    \label{fig:fpdemo}
\end{figure*}

Exoplanet candidates require vetting, in order to establish whether they are true planets or False Positives, (for example from blended background eclipsing binaries.) The \kepler pipeline, and tools such as the Robovetter \citep{robovetter}, employed some pixel level vetting of exoplanet candidates, to aid in distinguishing true planet candidates from False Positives for the \kepler mission. This vetting included analyzing centroid offsets to identify blended targets (see \cite{KDPH_DV1} and \cite{DR25}). The model presented here can be used in conjunction with existing tools to vet exoplanet candidates, and provide additional vetting statistics at the pixel level.

Unlike eclipsing binaries, exoplanet systems do not usually exhibit strong color dependence at visible wavelengths. Exoplanets usually do not emit in the visible, unless they have extremely high effective temperatures. As such, for true planets, we expect no transit depth changes due to color-dependence to be detectable using the techniques described in this work. We suggest that, using the model described in this work, it is possible to vet planet candidates by searching for significant shape variation in PRF during transit. Significant PRF shape changes would indicate the candidate planet is in fact a grazing eclipsing binary, with a significant color variation. Using chromaticity of the \kepler PRF to vet planet candidates can help distinguish between grazing eclisping binaries and grazing planets, which are otherwise difficult to distinguish (as both exhibit v-shaped transits).

Figure ~\ref{fig:fpdemo} shows an example of how the \kepler PRF chromaticity can be used to vet planet transit candidates, by comparing an eclipsing binary signal to a true planet candidate. Figure ~\ref{fig:fpdemo} shows the folded light curves of true planet Kepler-1b and eclipsing binary KIC 8703887, which shows an eclipse depth consistent with a large planet. These targets have been chosen as a demonstration owing to their similar transit/eclipse depths, similar durations, and similar apparent magnitudes. In Figure \ref{fig:fpdemo} we show the reduced chi squared metric of our PRF model fit for each cadence, which shows a significant disagreement during the eclipse of KIC 8703887. Kepler-1b shows no significant disagreement in reduced chi squared during transit, showing that it is not a False Positive eclipsing binary. Using the techniques presented in this work we can identify planet candidate targets where there is significant PRF shape variation during transit, and discount them as False positives. We note that a high signal to noise transit/eclipse would be required to use this technique.

There are some caveats when using this method to vet exoplanet transit signals in the \kepler dataset. Firstly, it is important to understand the magnitude of the PRF shape change due to wavelength-dependent variability as a function of \kepler magnitude and location on the detector. In this work we do not identify the magnitude limit of this effect. Secondly, it is important to account for astrophysical effects, such as emission for hot planets and wavelength dependent limb darkening. Accounting for such effects is beyond the scope of this work, but is enabled by the full \kepler dataset.

\subsubsection{Characterization: Flares}

Stellar flares have been studied extensively in \kepler data \citep[see e.g.][]{flares1, flares2}. However flare studies with \kepler have been adversely affected by two key factors; 1) \kepler's monochromatic band pass makes estimates of flare energy degenerate, as there is no way to constrain the flare temperature 2) the majority of \kepler data was taken in long-cadence mode, where each exposure is 30 minutes, which can make identification of short term flares on time scales $\lesssim$1 hour difficult. We propose that the chromaticity of the \kepler PRF can help alleviate both of these problems.

\begin{figure*}
    \centering
    \includegraphics[width=\textwidth]{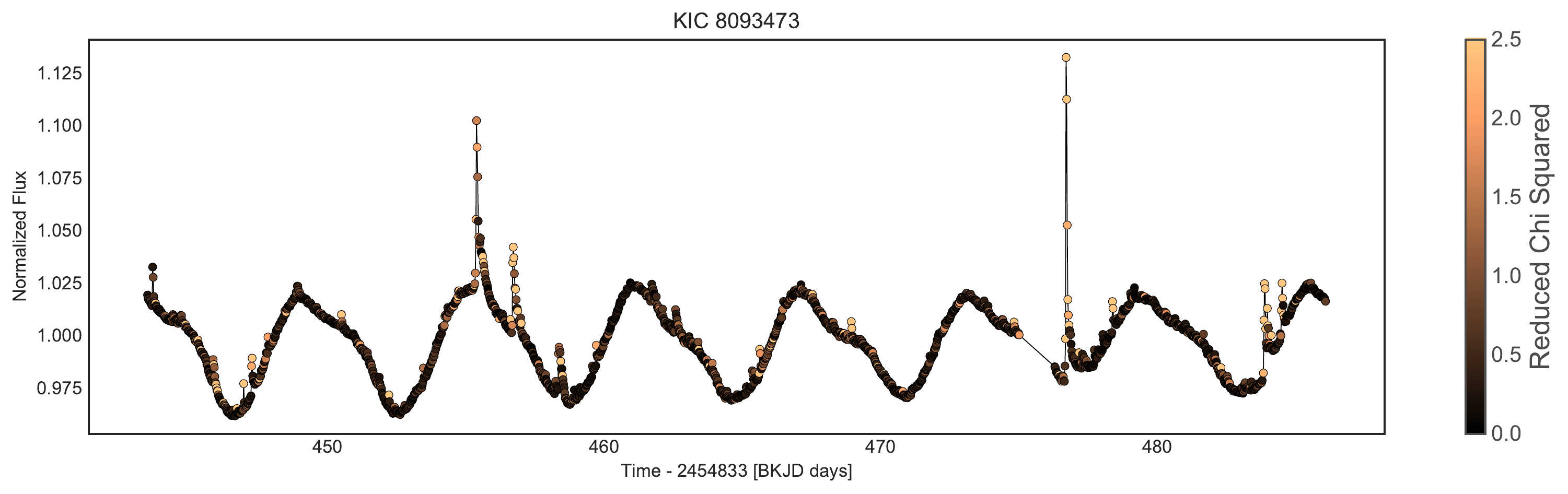}
    \caption{Demonstration of how the model presented in this work can be used to indentify and characterize flares. Shown is flare star KIC 8093473. The SAP flux light curve is plotted, with the color bar shows the reduced chi squared statistic at each point for our model compared with the measured \kepler data, summed up over all pixels. Several flare shows a significant deviation from the model, indicating a shape change. This is expected, given that stellar flares peak at a high temperature compared with the photosphere, and so have a wavelength dependence.}
    \label{fig:flare_demo}
\end{figure*}

Figure \ref{fig:flare_demo} shows an example light curve of a flaring \kepler target, with the reduced chi squared statistic of our systematics model. Flares are a significantly poorer fit than the stellar variability, showing that there is a change in the PRF shape during flares. We attribute this to a wavelength change during flares, as the flares are much hotter than the stellar photosphere we anticipate the target flux to be bluer during flares. Using the PRF shape change, it is possible to intercompare flares within a single target and investigate whether flares are all the same temperature (peak wavelength) or whether they vary.

Short term flares on timescales of $\lesssim$ 1 hour are frequently discarded by flare searches as they are indistinguishable from the signal of be cosmic rays. Using the model described here it would be possible to identify true flares by using the shape of the PRF, and rejecting cadences where the PRF shape is more consistent with a cosmic ray than with the target PRF, or PRF shape change due to color variability. Changes in shape due to flares should be centered on the main target, and should have a shape similar to a PSF, where as cosmic rays can appear in any pixel, and are usually either single pixels events or sharp, bright lines. Using simple image statistics (e.g. centroiding) it would be possible to distinguish these processes. This pixel level shape modeling could be capitalized on by machine learning algorithms, alongside time variable flux information, in order to better identify short time flares in the \kepler long cadence data set.

If a fully calibrated model for PRF shape were available, the wavelength dependence of the \kepler PRF could be used to extract temperature information on flares in the \kepler dataset. Currently we do not know the parameter space that would be probed by a fully calibrated model (e.g. completeness in \kepler magnitude, flare brightness, flare temperature e.t.c), however in theory it will be possible to at least place limits on the flare temperature, and to intercompare different flare energies within a single target.

\subsubsection{Characterization: Pulsators}

\begin{figure}
    \centering
    \includegraphics[width=0.45\textwidth]{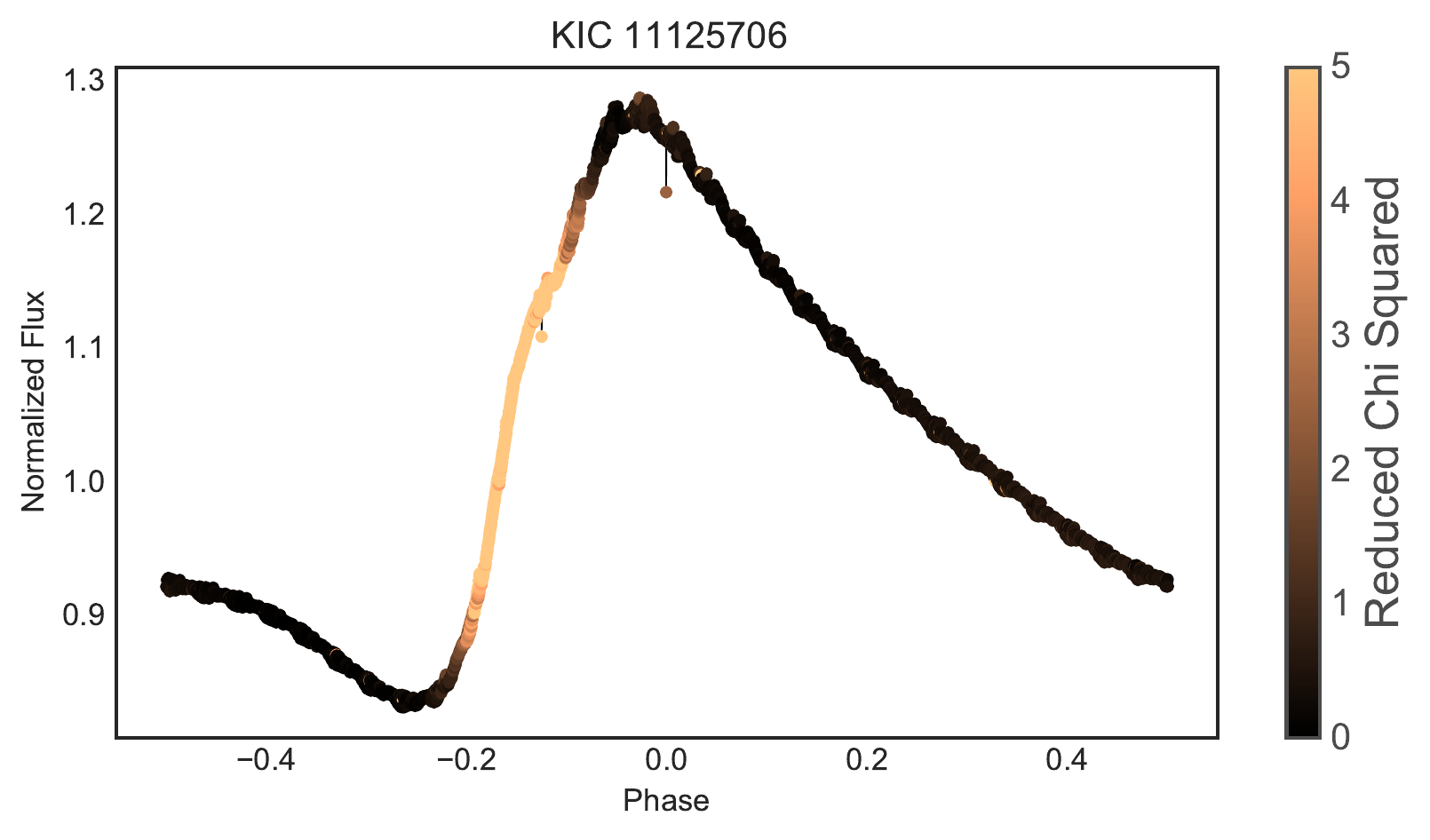}
    \caption{Demonstration of how the model presented in this work can be used to identify PSF shape changes for pulsating stars. Shown is KIC 11125706, an RR Lyrae star observed during the \kepler prime mission. Shown is the SAP flux light curve, folded at the period of the pulsations. Color bar shows the reduced chi squared statistic at each point for our model compared with the measured \kepler data, summed up over all pixels. There is a significantly poorer fit during the rapid brightening of the pulsator, indicating a change in color of the target at that phase.}
    \label{fig:pulsator_demos}
\end{figure}

Pulsating stars such as RR Lyrae are rarely studied at the precision and cadence of the \kepler spacecraft while also obtaining multiwavelength photometry. The chromatic PRF of \kepler provides a unique opportunity to study an enormous catalog of pulsators with a 4 year long observational baseline and high precision photometry, and investigate the wavelength dependence of their pulsations. Figure \ref{fig:pulsator_demos} shows an example light curve of an RR Lyrae star observed by \kepler, with the reduced chi-squared statistic or our pixel-level model fit at each phase of the pulsation. The fit is significantly poorer during the rapid rise of the pulsation, indicating that there may be a significant shape change in the PRF, and therefore a color dependence, during the pulsation. Better characterizing this color change across the sample of pulsators identified in \kepler, alongside \emph{Gaia} distances, may allow more robust characterization of the temperatures of these pulsators. 

\subsubsection{Characterization: Eclipsing Binaries}
As demonstrated in this work, eclipsing binaries are some of the best targets to see the effect of the chromatic \kepler PSF. Eclipse depths, phase curve variations and limb darkening are all variable across different wavelength. Even coarse multi-wavelength photometry can enable better understanding of stellar atmospheres, when coupled with the high cadence and long baseline time series photometry of \kepler. By employing a large sample of bright eclipsing binaries, and stellar spectra to obtain temperature estimates for each component, it would be possible to use this effect to understand the wavelength dependence of reflected light variation and ellipsoidal variation in eclipsing binaries, which in turn would teach us more about their outer atmospheres. It would also be possible to better calibrate limb darkening laws as a function of temperature and wavelength. Finally, with multiwavelength information, it may be possible model surface features such as spots with a wavelength dependent component, helping to constrain spot temperatures.

\subsection{Wavelength Dependence in Other Instruments and Data Sets}
The effect discussed in this work is not specific only to the \kepler spacecraft. The chromaticity of the \kepler PRF is due to wavelength dependence in the optical system, namely in the PSF as it is transmitted through the optical path of the telescope, and in the detector as the PSF is recorded. Other spacecraft will have analogous effects due to wavelength dependence in their optical systems. In particular, we highlight that the \tess spacecraft is anticipated to have a similar wavelength-dependence, owing to its use of lenses and large detector pixels. In fact, the magnitude of the effect for the \tess spacecraft is anticipated to be greater than that of \kepler. We encourage the investigation of other detectors to identify wavelength-dependent PSFs, which could unlock serendipitous wavelength-dependent photometry from other NASA time-series missions.

\subsubsection{Modifying the model for K2}
\label{sec:ktwo}
The model discussed in this work uses a simple polynomials in PRF centroid position to build our design matrix and model systematics. This is a reasonable assumption, as changes in the PRF centroid are gradual, (owing to focus change and velocity aberration) and are well modeled by a low order polynomial. In order to model data from the \ktwo phase of the \kepler mission, this model needs to be updated. \ktwo data exhibits a larger, high frequency PSF motion due to the roll of the spacecraft. This motion noise can be accounted for in our model by using the same method outlined in \cite{Vanderburg2014}, known as the Self Flat Fielding method, where centroids are converted into 'arclength' (distance along the motion vector in the imaging plane), and broken up in to short ~5-10 day bins. Our model can be modified to fit K2 data by removing the simple polynomials in centroid position and flux, and replacing it with a model for arclength and flux that is split into several 5-10 day bins, in order to remove the characteristic \ktwo motion noise.

\section{Summary}

The \kepler telescope and detector have a weak wavelength-dependent response due to chromatic aberration in the optics of the telescope and due to a wavelength dependent intrapixel sensitivity in the detector. These effects combine to give a weakly wavelength-dependent PRF (Pixel Response Function), which causes a shape change in the PRF depending on the color of the target. For bright variable stars that have different flux time-series at different wavelengths, this shape change can be measured in the \kepler data set.

In this work we have presented a new data-driven model for the \kepler PRF of a target, including instrument systematics, which models the effects of motion, velocity aberration and focus change. This model does not account for small variation in PRF shape due to wavelength dependence, and so wavelength dependent effects can be identified in the residuals between the data and our PRF model.

This work shows that for bright eclipsing binaries it is possible to extract coarse, multi-wavelength photometry across a range of $\approx$ 100nm. This has enabled us to reveal the wavelength dependence of phase curves, limb darkening and eclipse depth for our small sample of eclipsing binaries.

In this work we have approximately calibrated the effect using literature temperature values for a small set of eclipsing binaries. This work can be improved by expanding this sample and using robust temperature estimates with uncertainties. Fully calibrating the wavelength dependence of the \kepler PRF will require a targeted effort to build a data-driven PRF model as a function of 1) target brightness 2) target position across the detector and 3) wavelength. While there are theoretical PRF models for the \kepler spacecraft, these do not provide the accuracy needed to resolve the wavelength dependent effect. Given the resource provided by \emph{Gaia} of a full catalog of positions in three visible color-bands, it may be possible to build a data-driven PRF model by combining \emph{Gaia} and \kepler data. However, further investigation is needed.

In this work we have highlighted several potential use cases for this effect, ranging from classifying exoplanet targets and flare events, to making inference using color variability for variables such as eclipsing binaries and pulsators. There are dozens of ways even coarse multi-wavelength photometry can be used to enhance the legacy of \kepler. Using the model described in this work, the community will be able to access this valuable information, which has been otherwise been locked in the archive.

The method of this work is to model and remove all instrument systematics using pixel-level modeling, and identify PRF shape changes on a pixel level. This method is general, and while other instruments may require a different model for pixel systematics, the methods in this work can be applied to search for wavelength dependence in the PRF's of other instruments beyond \kepler. In particular, we highlight that the \tess spacecraft observes in a similar way to the \kepler spacecraft. Due to its instrument design of using lenses and larger pixels than \kepler, \tess is expected to have an even greater wavelength dependence than \kepler. We encourage the use of the model and techniques discussed here to search for chromatic effects in the data from other spacecraft.

\section{Acknowledgements}

This work has benefited greatly from the input of several members of the \kepler team, in particular we would like to acknowledge the contributions from Douglas Caldwell and Steven Bryson, who provided excellent consultations on the PRF for \kepler.

This paper includes data collected by the Kepler mission and obtained from the MAST data archive at the Space Telescope Science Institute (STScI). Funding for the Kepler mission is provided by the NASA Science Mission Directorate. STScI is operated by the Association of Universities for Research in Astronomy, Inc., under NASA contract NAS 5–26555. This research has made use of the NASA Exoplanet Archive, which is operated by the California Institute of Technology, under contract with the National Aeronautics and Space Administration under the Exoplanet Exploration Program. This research made use of Astropy,\footnote{http://www.astropy.org} a community-developed core Python package for Astronomy \citep{astropy2013, astropy2018}. This research made use of Lightkurve, a Python package for Kepler and TESS data analysis (Lightkurve Collaboration, 2018). We would like to thank the reviewer of this manuscript for their excellent review, which greatly benefited this publication.

\appendix
\begin{sidewaystable*}
\centering

\begin{tabular}{rrrrrrlllll}
\toprule
 KIC ID &  Kepler Mag &  T$_{eff1}^a$ &  T$_{eff2}^a$ &    Period &          t0 &                         Radius$_1$ &                         Radius$_2$ &                         Inclination &                      Amplitude$_1$ &                      Amplitude$_2$ \\
 
   &  &  $K$ &  $K$ &   days &   &                         $R_\odot$ &                         $R_\odot$ &                         $^{\circ}$ &                  &                       \\
\midrule
   2708156 &            10.672 &            11061 &             5671 &  1.891269 &  170.508742 &  2.27939$\pm_{-0.00011}^{0.00011}$ &  2.08991$\pm_{-0.00003}^{0.00003}$ &  82.61361$\pm_{-0.00045}^{0.00045}$ &  0.90881$\pm_{-0.00001}^{0.00001}$ &  0.09119$\pm_{-0.00001}^{0.00001}$ \\
   3241619 &            12.524 &             5715 &             4285 &  1.703347 &  169.942399 &  1.06836$\pm_{-0.00021}^{0.00021}$ &  1.02367$\pm_{-0.00018}^{0.00017}$ &  85.17719$\pm_{-0.00136}^{0.00139}$ &  0.76921$\pm_{-0.00014}^{0.00014}$ &  0.23079$\pm_{-0.00014}^{0.00014}$ \\
   4574310 &            13.242 &             7153 &             4077 &  1.306220 &  169.993896 &  1.49917$\pm_{-0.00043}^{0.00044}$ &  1.58066$\pm_{-0.00062}^{0.00059}$ &  78.49502$\pm_{-0.00672}^{0.00689}$ &  0.80368$\pm_{-0.00017}^{0.00017}$ &  0.19632$\pm_{-0.00017}^{0.00017}$ \\
   4665989 &            13.016 &             7559 &             6846 &  2.248068 &  170.972872 &  1.84602$\pm_{-0.00036}^{0.00037}$ &  1.36237$\pm_{-0.00086}^{0.00083}$ &  81.94781$\pm_{-0.00561}^{0.00567}$ &  0.74771$\pm_{-0.00030}^{0.00031}$ &  0.25229$\pm_{-0.00031}^{0.00030}$ \\
   4848423 &            11.825 &             6239 &             6176 &  3.003646 &  633.146586 &  1.50426$\pm_{-0.00023}^{0.00023}$ &  1.38874$\pm_{-0.00025}^{0.00026}$ &  87.57999$\pm_{-0.00069}^{0.00068}$ &  0.54403$\pm_{-0.00017}^{0.00016}$ &  0.45597$\pm_{-0.00016}^{0.00017}$ \\
   5444392 &            11.378 &             5965 &             5726 &  1.519529 &  170.538898 &  1.59339$\pm_{-0.00013}^{0.00013}$ &  1.66385$\pm_{-0.00009}^{0.00009}$ &  84.27161$\pm_{-0.00062}^{0.00062}$ &  0.51747$\pm_{-0.00007}^{0.00007}$ &  0.48253$\pm_{-0.00007}^{0.00007}$ \\
   5513861 &            11.638 &             6479 &             6411 &  1.510208 &  170.329100 &  1.64494$\pm_{-0.00040}^{0.00040}$ &  1.49045$\pm_{-0.00049}^{0.00047}$ &  78.72670$\pm_{-0.00234}^{0.00242}$ &  0.56207$\pm_{-0.00026}^{0.00027}$ &  0.43793$\pm_{-0.00027}^{0.00026}$ \\
   5738698 &            11.941 &             6792 &             6773 &  4.808774 &  171.679679 &  1.69030$\pm_{-0.00052}^{0.00053}$ &  1.53567$\pm_{-0.00067}^{0.00067}$ &  86.19971$\pm_{-0.00154}^{0.00158}$ &  0.55585$\pm_{-0.00036}^{0.00037}$ &  0.44415$\pm_{-0.00037}^{0.00036}$ \\
   6206751 &            12.142 &             6965 &             4885 &  1.245343 &  170.824532 &  1.50527$\pm_{-0.00164}^{0.00160}$ &  1.58397$\pm_{-0.00308}^{0.00309}$ &  71.73854$\pm_{-0.02101}^{0.02087}$ &  0.77538$\pm_{-0.00115}^{0.00115}$ &  0.22462$\pm_{-0.00115}^{0.00115}$ \\
   8262223 &            12.146 &             9128 &             6849 &  1.613015 &  170.460863 &  1.36443$\pm_{-0.00310}^{0.00301}$ &  1.82295$\pm_{-0.00338}^{0.00345}$ &  72.16185$\pm_{-0.00712}^{0.00714}$ &  0.58017$\pm_{-0.00229}^{0.00226}$ &  0.41983$\pm_{-0.00226}^{0.00229}$ \\
   8553788 &            12.691 &             8045 &             5328 &  1.606164 &  170.180393 &  1.95352$\pm_{-0.00604}^{0.00600}$ &  1.61936$\pm_{-0.01344}^{0.01344}$ &  70.23075$\pm_{-0.09297}^{0.09245}$ &  0.91970$\pm_{-0.00184}^{0.00182}$ &  0.08030$\pm_{-0.00182}^{0.00184}$ \\
   8823397 &            13.249 &             8540 &             5724 &  1.506504 &  170.654968 &  1.69962$\pm_{-0.00016}^{0.00017}$ &  1.25994$\pm_{-0.00027}^{0.00027}$ &  84.49796$\pm_{-0.00570}^{0.00578}$ &  0.85919$\pm_{-0.00006}^{0.00006}$ &  0.14081$\pm_{-0.00006}^{0.00006}$ \\
   9159301 &            12.146 &             7959 &             4209 &  3.044772 &  172.019893 &  2.85734$\pm_{-0.00003}^{0.00003}$ &  1.99571$\pm_{-0.00003}^{0.00003}$ &  85.84383$\pm_{-0.00047}^{0.00049}$ &  0.95374$\pm_{-0.00001}^{0.00001}$ &  0.04626$\pm_{-0.00001}^{0.00001}$ \\
   9357275 &            12.186 &             7545 &             5580 &  1.588298 &  170.303369 &  1.55017$\pm_{-0.00277}^{0.00279}$ &  1.60682$\pm_{-0.00490}^{0.00485}$ &  72.45522$\pm_{-0.02677}^{0.02667}$ &  0.80551$\pm_{-0.00171}^{0.00173}$ &  0.19449$\pm_{-0.00173}^{0.00171}$ \\
   9592855 &            12.216 &             7290 &             7087 &  1.219325 &  907.436255 &  1.59043$\pm_{-0.04540}^{0.04555}$ &  1.63729$\pm_{-0.04631}^{0.04601}$ &  68.48705$\pm_{-0.02219}^{0.02090}$ &  0.49956$\pm_{-0.03212}^{0.03222}$ &  0.50044$\pm_{-0.03222}^{0.03212}$ \\
   9602595 &            11.882 &             9679 &             5705 &  3.556513 &  172.643854 &  2.59819$\pm_{-0.00017}^{0.00017}$ &  3.09062$\pm_{-0.00006}^{0.00006}$ &  81.85038$\pm_{-0.00044}^{0.00045}$ &  0.88844$\pm_{-0.00002}^{0.00002}$ &  0.11156$\pm_{-0.00002}^{0.00002}$ \\
   9851944 &            11.249 &             7026 &             6902 &  2.163902 &  170.471968 &  1.81528$\pm_{-0.00102}^{0.00099}$ &  2.72157$\pm_{-0.00084}^{0.00086}$ &  73.91465$\pm_{-0.00370}^{0.00374}$ &  0.31987$\pm_{-0.00040}^{0.00039}$ &  0.68013$\pm_{-0.00039}^{0.00040}$ \\
   9899416 &            10.028 &            11056 &             6278 &  1.332564 &  170.693626 &  2.24077$\pm_{-0.00122}^{0.00113}$ &  1.73692$\pm_{-0.00101}^{0.00103}$ &  87.46126$\pm_{-0.13032}^{0.13031}$ &  0.88206$\pm_{-0.00006}^{0.00006}$ &  0.11794$\pm_{-0.00006}^{0.00006}$ \\
  10156064 &            10.367 &             7424 &             6268 &  4.855936 &  170.017881 &  1.36860$\pm_{-0.00157}^{0.00150}$ &  1.79987$\pm_{-0.00161}^{0.00170}$ &  81.72808$\pm_{-0.00122}^{0.00118}$ &  0.55287$\pm_{-0.00117}^{0.00111}$ &  0.44713$\pm_{-0.00111}^{0.00117}$ \\
  10191056 &            10.811 &             6588 &             6455 &  2.427495 &  170.581538 &  1.84616$\pm_{-0.00072}^{0.00073}$ &  1.63960$\pm_{-0.00083}^{0.00080}$ &  78.92767$\pm_{-0.00193}^{0.00189}$ &  0.56878$\pm_{-0.00045}^{0.00045}$ &  0.43122$\pm_{-0.00045}^{0.00045}$ \\
  10486425 &            12.465 &             7018 &             5847 &  5.274818 &  173.633819 &  1.68109$\pm_{-0.00132}^{0.00128}$ &  0.93051$\pm_{-0.00426}^{0.00432}$ &  83.97970$\pm_{-0.01582}^{0.01546}$ &  0.93073$\pm_{-0.00075}^{0.00072}$ &  0.06927$\pm_{-0.00072}^{0.00075}$ \\
  10619109 &            11.704 &             7028 &             3903 &  2.045166 &  171.224091 &  1.78472$\pm_{-0.00175}^{0.00177}$ &  2.45157$\pm_{-0.00357}^{0.00352}$ &  70.46071$\pm_{-0.01975}^{0.01999}$ &  0.84881$\pm_{-0.00066}^{0.00067}$ &  0.15119$\pm_{-0.00067}^{0.00066}$ \\
  10661783 &             9.586 &             7764 &             6001 &  1.231363 &  169.978747 &  0.86273$\pm_{-0.00001}^{0.00001}$ &  2.79316$\pm_{-0.00002}^{0.00002}$ &  72.36971$\pm_{-0.00030}^{0.00030}$ &  0.18305$\pm_{-0.00000}^{0.00000}$ &  0.81695$\pm_{-0.00000}^{0.00000}$ \\
  10686876 &            11.727 &             7944 &             5842 &  2.618415 &  170.700071 &  0.57829$\pm_{-0.00006}^{0.00006}$ &  2.80106$\pm_{-0.00014}^{0.00014}$ &  76.28074$\pm_{-0.00118}^{0.00115}$ &  0.21066$\pm_{-0.00002}^{0.00002}$ &  0.78934$\pm_{-0.00002}^{0.00002}$ \\
  10736223 &            13.621 &             7797 &             5069 &  1.105094 &  170.119937 &  1.49741$\pm_{-0.00008}^{0.00008}$ &  1.29056$\pm_{-0.00004}^{0.00004}$ &  88.43092$\pm_{-0.00205}^{0.00207}$ &  0.88487$\pm_{-0.00002}^{0.00002}$ &  0.11513$\pm_{-0.00002}^{0.00002}$ \\
  10858720 &            10.971 &             7282 &             7223 &  0.952378 &  170.430133 &  1.54288$\pm_{-0.00001}^{0.00001}$ &  1.36804$\pm_{-0.00001}^{0.00001}$ &  88.33953$\pm_{-0.00047}^{0.00047}$ &  0.57041$\pm_{-0.00000}^{0.00000}$ &  0.42959$\pm_{-0.00000}^{0.00000}$ \\
\bottomrule
\end{tabular}
\label{tab:parameters}
\caption{Table of parameters of the eclipsing binaries used in this work. The \kepler magnitudes and effective temperatures have been derived from \cite{ebliterature}. Here subscript of $1$ denotes the primary body, and subscript of $2$ denotes the secondary. The radius, inclination and amplitude values were derived in this work by fitting an eclipsing binary model using \texttt{starry} \citep{starry}.}
\footnotesize{$^a$ Effective temperatures are from \cite{ebliterature}}
\end{sidewaystable*}

\bibliographystyle{aasjournal}
\bibliography{bib.bib}

\end{document}